\documentclass[letterpaper,11pt]{article}
\pdfoutput=1

\usepackage{jheppub}
\usepackage{multirow}
\usepackage{subfig}
\usepackage{xspace}
\usepackage{xcolor}
\usepackage[countmax]{subfloat}
\usepackage{amsmath}
\usepackage{mathtools}
\usepackage{graphicx}
\usepackage{slashed}
\usepackage{tikz-feynman,contour}
\usepackage{tikz,pgf}
\usepackage{braket}
\usepackage{amsfonts}
\usepackage{amsthm,scrextend,bbold}
\usepackage{comment}

\newcommand{\as}{\alpha_\text{s}}
\newcommand{\cf}{C_{\text{F}}}
\newcommand{\ca}{C_{\text{A}}}

\newcommand{\msb}{\overline{\text{MS}}}
\newcommand{\order}[1]{{\cal O}\left(#1\right)}

\DeclareMathOperator{\De}{d}
\newcommand{\de}{\De\!}
\newcommand{\xf}{x}

\newcommand{\gs}{\gamma_\text{soft}}
\newcommand{\gszero}{\gamma_\text{soft}^{(0)}}

\newcommand{\gszerotilde}{\widetilde{\gamma}_\text{soft}^{(0,\text{sub})}}

\newcommand{\muf}{\mu_{\text{F}}}
\newcommand{\muOf}{\mu_{0\text{F}}}
\newcommand{\mur}{\mu_{\text{R}}}
\newcommand{\muOr}{\mu_{0\text{R}}}

\usepackage{color}
\definecolor{darkblue}{rgb}{0,0,0.5}
\definecolor{darkgreen}{rgb}{0,0.5,0}
\definecolor{darkorange}{rgb}{0.8,0.3,0}

\title{
A consistent resummation of mass and soft logarithms in processes with heavy flavours
}

\author[1]{Andrea Ghira,}
\author[1]{Simone Marzani,}
\author[1]{and Giovanni Ridolfi}

\affiliation[1]{Dipartimento di Fisica, Universit\`a di Genova and INFN, Sezione di Genova,Via Dodecaneso 33, 16146, Italy}
\emailAdd{andrea.ghira@ge.infn.it}
\emailAdd{simone.marzani@ge.infn.it}
\emailAdd{giovanni.ridolfi@ge.infn.it}

\abstract{
Perturbative calculations for processes that involve heavy flavours can be performed in two approaches: the massive scheme and the massless one. The former enables one to fully account for the heavy-quark kinematics, while the latter allows one to resum potentially-large mass logarithms.
Furthermore, the two schemes can be combined to take advantage of the virtues of each of them.
Both massive and massless calculations can be supplemented by soft-gluon resummation. 
However matching between massive and massless resummed calculations is difficult, essentially because of the non-commutativity of the soft and massless limits. 
In this paper, we develop a formalism to combine resummed massive and massless calculations. We obtain an all-order expression that consistently resums both mass and soft logarithms to next-to-leading logarithmic accuracy. We perform detailed calculations for the decay of the Higgs into a heavy-quark pair, and discuss the applications of this formalism to different processes. 
}
\begin{document}
\maketitle

\section{Introduction}\label{sec:intro}

The physics of heavy flavours is especially important in particle physics phenomenology, for a number of reasons.
The Higgs boson decays primarily into pairs of $b$ quarks and, although this decay mode is challenging because of its large background, it plays a central role in studies of electro-weak symmetry breaking. 
Furthermore, many aspects of so-called flavour-physics can be scrutinised in the charm and bottom sectors.

Heavy flavours are also a valuable probe of strong interactions. Despite the fact that gluons couple to quarks irrespectively of their mass, quark masses do affect emergent phenomena, such as jet formation and their substructure. 
In this context, a noteworthy effect is the so-called dead-cone, i.e.\ the fact that colour radiation around heavy quarks is suppressed~\cite{Dokshitzer:1991fd,Dokshitzer:1995ev}. 
Dedicated phenomenological strategies have been designed to expose and study this effect, e.g.~\cite{Llorente:2014bha,Maltoni:2016ays,Cunqueiro:2018jbh}, which has been recently measured by the ALICE collaboration at the LHC~\cite{ALICE:2021aqk}. Furthermore, the possibility of exploiting the imprint that quark masses leave on colour correlations has been recently investigated in the context of $b$-tagging~\cite{Fedkevych:2022mid}.
Moreover, while the top quark mass is so large that this particle's lifetime is shorter than the typical time scale of hadron formation, 
the mass of $c$ and $b$, although in the perturbative regime, is not so large so that hadron formation occurs.
Thus, processes involving $c$ and $b$ quarks can be exploited to scrutinise the mechanism that binds quarks and gluons into colour-neutral hadrons. 
Finally, studies of intrinsic heavy-flavour component (mostly charm quarks) in the proton requires precision calculations of perturbative cross-section involving heavy quarks.  

Two main strategies to perform QCD calculations of observables with heavy flavours are usually employed. In  the so-called \emph{massive scheme}, heavy quarks in the final state are considered as real, on-shell, particles with a non-zero mass.
The main advantage of the massive scheme is that the kinematics of the produced heavy flavour is treated exactly, because the full mass-dependence is retained. 
An important example of a calculation performed in this scheme is 
heavy flavour production in hadron-hadron collisions, which has been computed up to next-to-next-to-leading order (NNLO), see e.g.~\cite{Czakon:2013goa,Catani:2019iny,Catani:2020kkl,Catani:2022mfv,Buonocore:2022pqq,Buonocore:2023ljm}
The fixed-order precision can be improved by various types of all-order calculations, e.g.\  soft-gluon-resummation, for both inclusive production cross sections and differential distributions~\cite{Cacciari:2011hy,Gaggero:2022hmv}, high-energy resummation~\cite{Catani:1990eg,Ball:2001pq} or even transverse momentum resummation~\cite{Catani:2014qha}.
The inclusion of heavy-quark effects in general-purpose Monte Carlo parton shower codes is also an area of active research, see e.g.~\cite{Norrbin:2000uu,Bahr:2008pv,Krauss:2017wmx,Assi:2023rbu}.

The range of energies probed by collider experiments is typically much larger than the heavy-flavour mass, making heavy-flavour production a multi-scale problem. Theoretical predictions for these processes, even for inclusive observables, are plagued by logarithms of $m^2/q^2$, where $q^2$ the square of the hard scale, that can spoil the convergence of the perturbative expansion. 
Therefore an alternative calculational framework is often employed. This second approach exploits \emph{fragmentation functions} to resum these logarithmic corrections to all orders. This is possible because these logarithmic corrections are related to collinear dynamics, which would give rise to divergencies in a massless theory. It follows that, up to corrections $\order{m^2/q^2}$, heavy-flavour production cross-sections obey a factorisation theorem and can be written as the convolution of process-dependent partonic (massless) coefficient functions with universal heavy-flavour fragmentation functions. Fragmentation functions obey DGLAP evolutions equations (with time-like splitting functions), which allow one to resum the large logarithmic corrections we are discussing, in analogy to the initial-state collinear factorisation theorem. 
The initial-condition for heavy quark fragmentation functions can be computed in perturbation theory, as originally pointed out  in Ref.~\cite{Mele:1990cw,Mele:1990yq}, where a NLO computation is presented.
The NNLO corrections were computed later in Refs.~\cite{Melnikov:2004bm,Mitov:2004du}. 
The initial condition of the evolution is, by definition, free of mass logarithms, but it is affected by soft logarithms, that can be resummed to all orders too.~\cite{Cacciari:2001cw,Fickinger:2016rfd,Maltoni:2022bpy,Czakon:2022pyz}

Virtues of the massive and massless schemes can be combined together by matching the fixed-order calculation performed in the massive scheme, with the all-order resummation of mass logarithms achieved by the fragmentation function approach~\cite{Cacciari:1998it,Cacciari:2001td,Forte:2010ta,Forte:2016sja,Bonvini:2015pxa,Pietrulewicz:2017gxc,Ridolfi:2019bch}.
Since soft-gluon resummation is available for both the massive and the massless schemes, it is natural to explore the possibility of matching the two calculations at the resummed level, obtaining a theoretical prediction that fully accounts for all mass and soft logarithms. 
As it has been pointed out in the literature, see e.g.~\cite{Cacciari:2002re,Corcella:2003ib,Mitov:2003bm,Gaggero:2022hmv,Aglietti:2007bp,Aglietti:2022rcm} this is far from trivial, because the structure of soft logarithms significantly differs in the two approaches, hampering the construction of an all-order matching scheme.
The goal of this paper is to overcome this difficulty and build a matching scheme that allows us to consistently resum both mass and soft logarithms in processes with heavy quarks. 
The crucial ingredient for the construction of a consistent resummation formula is a modification of the standard fragmentation-function calculation that, by exploiting the so-called quasi-collinear limit~\cite{Catani:2000ef,Catani:2002hc}, correctly accounts for mass effects that originate from QCD radiation from all (hard) heavy-quarks in the process.  
Furthermore, we will also discuss in some details the inclusion of heavy-quark threshold effects that arise from the treatment of the QCD running coupling in a decoupling scheme, which is more natural than standard $\msb$ when considering processes with heavy flavours.
In this work, we concentrate on parton-level results, leaving detailed phenomenological analyses to future work. 

In our discussion, we are going to focus on the decay of an (off-shell) electroweak boson into two massive quarks, considering  $H \to b \bar b$ as a concrete example. However, because our construction is general, we will briefly discuss how to apply it to other processes involving heavy flavours, such as deep-inelastic scattering and heavy-quark decay. For the latter, we will also comment on similarities and differences between our calculation and an alternative approach developed in Refs.~\cite{Aglietti:2007bp,Aglietti:2022rcm}.

This paper is organised as follows. In Sect.~\ref{sec:decay} we review known results on soft-gluon resummation in the massive and massless schemes, highlighting the issues that we are set to solve. In Sect.~\ref{sec:jet_mom} we revisit the problem in momentum space. This will allow us to better identify the relevant kinematic regions, leading to a more refined calculation. We will present our results for $H\to b \bar b$ in Sect.~\ref{sec:results_decay}, comment on other processes and approaches in Sect.~\ref{sec:extensions}, before drawing our conclusions in Sect.~\ref{sec:conclusions}. Details of the calculations and explicit results are collected in the appendices. 

\section{Heavy flavour pair production in weak decays}\label{sec:decay}
In this section we consider the decay of a colour-singlet massive system $H$, for instance an off-shell photon, a Higgs boson or a $Z$ boson, into a heavy quark-antiquark pair
plus undetected radiation:
\begin{equation}
	H(q)\to b(p_1)+\bar{b}(p_2)+X(k),
	\label{eq:Hdecay}
\end{equation}
(four-momenta are indicated in brackets.) In what follows, it is understood that the heavy quark is a $b$ quark, but similar considerations can be extended to charm.
We are interested in the differential decay rate $\frac{\de \Gamma}{\de x}$ with respect to the dimensionless variable
\begin{equation}
	x=\frac{2 p_1\cdot q}{q^2},
	\label{eq:xdef}
\end{equation}
which coincides with the fraction of the total available energy carried away by the $b$ quark in the centre of mass frame. At lowest order in QCD perturbation theory, when no radiation is present, energy-momentum conservation gives $x=1$:
\begin{equation}
\frac{\de\Gamma}{\de x}=\Gamma_0\delta(1-x)+\mathcal{O}(\as).
\end{equation}
At higher orders the $x$ dependence of the differential rate becomes non-trivial, and the limit $x\to1$ corresponds to kinematical configurations in which the emitted radiation is either soft, or collinear to the final-state antiquark.

The calculation of the differential rate can be carried out in two different schemes, sometimes referred to as the massive (or 4-flavour) and the massless (or 5-flavour) schemes.
In the first approach, the spectrum $\frac{\de \Gamma}{\de x}$ is computed to some finite order in perturbation theory taking into account the finite value of the heavy quark mass $m$ exactly. In this approach, collinear singularities are regularised by the heavy quark mass and,  consequently,
powers of  $\log\frac{m^2}{q^2}$ appear in the perturbative coefficients; such logarithms are large as $q^2\gg m^2$, and may eventually spoil the convergence of the perturbative series. On the other hand, the kinematics of radiation is exactly taken into account in the whole range, up to a given order in 
$\as$.
In the second approach, all flavours, including the $b$ quark, are treated as massless, which is an accurate approximation as long as $q^2\gg m^2$. Heavy-quark production is described in the fragmentation function formalism: factorisation is exploited in the sense that each final-state parton can fragment into a heavy quark with a suitable probability distribution, in close analogy to what happens for initial-state parton distribution functions. The mechanism of factorisation of collinear singularities is also very similar.
In this approach, the heavy flavour mass is retained only as a regulator of collinear divergences, while contributions proportional to powers of $\frac{m^2}{q^2}$ are systematically neglected. 
Within this framework, the differential decay rate takes the factorized form
	\begin{align}\label{eq:rate-frag}
		\frac{\de\Gamma}{\de\xf}&= \Gamma_0 \sum_{i} \int_x^1 \frac{\de z}{z} \mathcal{C}_i\left(\frac{x}{z},\frac{\muf^2}{q^2},\frac{\mur^2}{q^2} ,\as(\mur^2) \right) \mathcal{D}_i (z,\muf^2,m^2) +\order{\frac{m^2}{q^2}},
	\end{align}
where $\muf^2$ and $\mur^2$ are the factorisation and renormalisation scales, typically chosen of the order of the hard scale $q^2$. The sum runs over all partons; the functions $\mathcal{C}_i$ are process-dependent partonic cross sections, which admit a perturbative expansion in powers of $\as$ as long as $q^2$ is large enough. They are convoluted with the process-independent fragmentation functions $\mathcal{D}_i$. In this approach, only the dominant collinear region of radiation is included (as opposed to the massive scheme, where large-angle radiation is perturbatively taken into account). On the other hand, collinear logarithms of $\frac{m^2}{q^2}$ are resummed to all orders, up to a given logarithmic accuracy, through the solution of DGLAP equations for the fragmentation functions. Furthermore, the initial condition for the evolution equation of the $b$ fragmentation function is given at an initial scale of the order of the $b$ mass, and can therefore can be computed perturbatively~\cite{Mele:1990cw,Cacciari:2001cw, Melnikov:2004bm}.

The convolution product in Eq.~(\ref{eq:rate-frag}) is turned into an ordinary product by Mellin transformation, defined as
\begin{equation}
\tilde f(N)=\int_0^1dx\,x^{N-1}f(x)
\end{equation}
for a generic function $f(x)$ defined in the range $0\le x\le 1$. We find
	 \begin{align}\label{eq:rate-frag-N}
		\widetilde{\Gamma}(N,\xi)&= \frac{1}{\Gamma_0}\int_0^1 \de x \, x^{N-1}\,  \frac{\de\Gamma}{\de\xf}
		=  \sum_{i}  \widetilde{\mathcal{C}}_i\left(N, \frac{\muf^2}{q^2} , \frac{\mur^2}{q^2},\as(\mur^2)\right) \widetilde{\mathcal{D}}_i (N,\muf^2,m^2) + \order{\xi},
	\end{align}
where we have defined $\xi=\frac{m^2}{q^2}$.
The scale dependence of the fragmentation functions is governed by the DGLAP evolution equations
\begin{align}\label{eq:dglap}
	\muf^2 \frac{\de }{\de \, \muf^2}\widetilde{\mathcal{D}}_i (N,\muf^2,m^2) = \sum_j \gamma_{ij}\left(N,\as(\muf^2) \right) \widetilde{\mathcal{D}}^j (N,\muf^2,m^2).
\end{align}
The anomalous dimensions $\gamma_{ij}$, Mellin transforms of the time-like splitting function, can be computed perturbatively; their explicit expressions to order $\as$
can be found e.g.\ in~\cite{Ellis:1996mzs}. The solution of Eq.~(\ref{eq:dglap}) can be schematically written in terms of a matrix of evolution kernels $\widetilde{\mathcal{E}}_{ij}$ and initial conditions, given at some reference scale $\muOf$:
\begin{equation}\label{eq:dglap-sol}
	\widetilde{\mathcal{D}}_i (N,\muf^2,m^2) = \sum_j \widetilde{\mathcal{E}}_{ij}(N,\muOf^2,\muf^2)  \, \widetilde{\mathcal{D}}_{0}\;^j (N, \muOf^2,m^2).
\end{equation} 

In the following, we will denote by $\widetilde{\Gamma}^{(4)}_k(N,\xi)$ the (Mellin transformed) spectrum computed in the massive scheme up to order $k$, and by 
$\widetilde{\Gamma}^{(5)}_\ell(N,\xi)$ the same quantity computed in the massless schemes, with evolution equations for the fragmentation functions solved at next$^\ell$-to-leading logarithmic accuracy, and both the coefficient functions $\widetilde{\mathcal{C}}_i$ and the initial conditions for the evolution of fragmentation functions $\widetilde{\mathcal{D}}_{0}\;^j $ computed up to order $k$.
One may take advantage of both calculation schemes by combining the two results:
\begin{equation}
	\widetilde{\Gamma}(N,\xi)= \widetilde{\Gamma}^{(4)}_k(N,\xi)+\widetilde{\Gamma}^{(5)}_\ell(N,\xi)-\text{double counting},
	\label{FONLL}
\end{equation}
where ``double counting" stands for the perturbative expansion of  $\widetilde{\Gamma}^{(5)}_\ell(N,\xi)$ to order $k$. In the following we will restrict ourselves to the case $\ell=1$, which  is referred to as the FONLL scheme~\cite{Cacciari:1998it}. 

The perturbative coefficients of both quantities appearing in the rhs of Eq.~(\ref{FONLL}) display a logarithmically divergent behaviour as $N\to\infty$.
Such logarithms arise as a remnant of the cancellation of soft singularities, which results in the presence of distributions
\begin{equation}
D_k(x)=\left[\frac{\log^{k-1}(1-x)}{1-x}\right]_+
\label{Dk}
\end{equation}
in the physical spectrum, which dominate the perturbative coefficients in the vicinity of the threshold region $x\to 1$. 
Mellin transformation maps the large-$x$ region into the large-$N$ region; it can be shown that at large $N$ the Mellin transform of $D_k(x)$ is a polynomial of degree $k$ in $\log N$. 
These logarithmic contributions can be summed to all orders in the QCD expansion, up to a given logarithmic accuracy.

The quantities appearing in Eq.~(\ref{FONLL}) have different structures at large $N$. The order-$\as^n$ perturbative coefficient in the expansion of $\widetilde{\Gamma}^{(4)}_k(N,\xi)$ is a polynomial of degree $n$ in $\log N$, plus terms that vanish as $N\to+\infty$. The coefficients of the polynomial carry the full $\xi$ dependence: they include both powers of $\xi$, that vanish in the massless limit, and powers of $\log\xi$, which regularize collinear singularities. The coefficient functions $\mathcal{C}_i(N,\muf^2/q^2,\mur^2/q^2,\as(\mur^2)$ in 
$\widetilde{\Gamma}^{(5)}_\ell(N,\xi)$ are ordinary, subtracted partonic cross sections in massless QCD: the perturbative coefficients contain up to two powers of $\log N$ for each power of $\as$, corresponding to the remnants of soft singularity cancellation and collinear singularity subtraction. Finally, the initial conditions  for the evolution of the fragmentation functions $\widetilde{\mathcal{D}}_{0}\;^j (N, \muOf^2,m^2)$ can be computed perturbatively; the corresponding order-$\as^n$ coefficients are polynomials of degree $2n$ in $\log N$, whose coefficients depend on $\log\xi$ but not on powers of $\xi$. 

As mentioned in the introduction, the resummation of soft-gluon contributions at all orders, up to a given logarithmic accuracy, for both $\widetilde{\Gamma}^{(4)}_k(N,\xi)$ and $\widetilde{\Gamma}^{(5)}_\ell(N,\xi)$ can be performed, and has been studied in the in the past. In particular, for the massive-scheme case, we have
\begin{equation}
	\widetilde{\Gamma}^{(4)}(N,\xi)= \widetilde{\Gamma}^{(4)}_k(N,\xi)+\widetilde{\Gamma}^{(4,\text{res})}_{\ell_1}(N,\xi)
		-\text{double counting}
	\label{eq:gamma4}
\end{equation}
where  $\widetilde{\Gamma}^{(4,\text{res})}_{\ell_1}(N,\xi)$ is the massive-scheme spectrum resummed to next$^{\ell_1}$-to-leading $\log N$, and the ``double counting" term is its expansion to order $k$. $\widetilde{\Gamma}^{(4,\text{res})}_{\ell_1}(N,\xi)$ was computed in~\cite{Gaggero:2022hmv} for $\ell_1=2$.
Similarly
\begin{equation}
	\widetilde{\Gamma}^{(5)}(N,\xi)= \widetilde{\Gamma}^{(5)}_\ell(N,\xi)
	+\widetilde{\Gamma}^{(5,\text{res})}_{\ell\ell_2}(N,\xi)
	-\text{double counting}
	\label{eq:gamma5}
\end{equation}
and $\widetilde{\Gamma}^{(5,\text{res})}_{\ell\ell_2}(N,\xi)$ is the massless-scheme spectrum with both the coefficient functions and the initial conditions resummed to next$^{\ell_2}$-to-leading $\log N$, and the ``double counting" term is its expansion to order $k$.

The combination of Eq.~(\ref{eq:gamma4}) and Eq.~(\ref{eq:gamma5}) in a single formula, which would incorporate the best of theoretical knowledge of the $b$ energy spectrum in perturbative QCD, is not straightforward. 
Indeed, the sum of the results in Eq.~(\ref{eq:gamma4}) and Eq.~(\ref{eq:gamma5}) would require a further subtraction of doubly-counted contributions at all orders in $\as$; such subtraction is far from trivial, because the two limits $N\to\infty$ and $\xi\to 0$ are known not to commute~\cite{Cacciari:2002re,Corcella:2003ib,Mitov:2003bm,Gaggero:2022hmv,Aglietti:2007bp,Aglietti:2022rcm}: one cannot simply subtract from the $\widetilde{\Gamma}^{(4,\text{res})}_{\ell_1}(N,\xi)$ term in Eq.~(\ref{eq:gamma4}) its expression in the $\xi\to 0$ limit, because it differs from the
$N\to\infty$ limit of $\widetilde{\Gamma}^{(5,\text{res})}_{\ell\ell_2}(N,\xi)$.

In the following, after reviewing the expressions for the resummation of threshold logarithms in both the massive and the massless scheme, we will present a solution to this problem, valid to next-to-leading logarithmic accuracy.

\subsection{Soft resummation in the massive scheme}
The calculation of $\widetilde{\Gamma}^{(4,\text{res})}_{\ell_1}(N,\xi)$ was performed in Ref.~\cite{Laenen:1998qw}. It takes the exponentiated form 
\begin{align}\label{eq:massive resummation}
	\widetilde{\Gamma}^{(4,\text{res})}_{\ell_1}(N,\xi)=\mathcal{K}(\xi,\as) e^{2\int^1_0 \de x \frac{x^{N-1}-1}{1-x} \gs \left(\xi,\as\left((1-x)^2 q^2\right)\right)},
\end{align}
where $\mathcal{K}(\xi,\as)$ is a process-dependent factor, and $\gs$, the so-called massive soft anomalous dimension, admits an expansion in powers of $\as$:
\begin{equation} \label{ref:gs0}
\gs(\xi,
\as(\mu^2))=\frac{\as(\mu^2)}{\pi}\gszero(\beta)+\mathcal{O}(\as^2),
\end{equation}
where
\begin{equation}
\beta=\sqrt{1-4\xi}.
\end{equation}
An explicit calculation gives
\begin{equation}
	\gszero(\beta)=\cf\left(\frac{1+\beta^2}{2\beta}\log\frac{1+\beta}{1-\beta}-1\right).
	\label{eq:gammazero}
\end{equation}
Note that the resummed expression in Eq.~(\ref{eq:massive resummation}) features, at most, single  logarithms of $N$ to any order in perturbation theory, i.e.\ $\as^n \log^m N$, with $m\le n$. This is not surprising: collinear singularities are absent because of the finite quark mass and, consequently, collinear logarithms do not appear as logarithms of $N$ but rather as logarithms of the mass.
Thus, knowledge of the massive soft anomalous dimension to  order $\ell_1+1$ allows us to perform soft resummation at N$^{\ell_1}$LL accuracy. The soft anomalous dimension is currently known to second order~\cite{Korchemsky:1987wg,Kidonakis:2009ev,Becher:2009kw,vonManteuffel:2014mva}. However, in this work we restrict ourselves to resummation to single-logarithmic accuracy, both in the massless and in the massive schemes, so we consider $\ell_1=0$, i.e.\ we take the massive soft anomalous dimension at one loop, and the running of the strong coupling in Eq.~(\ref{eq:massive resummation}) is taken into account at the leading logarithmic level.

At this accuracy, the Mellin transform Eq.~(\ref{eq:massive resummation}) can be performed~\cite{Catani:1989ne} by the replacement 
\begin{equation}\label{eq:large-N integral}
	x^{N-1}-1\to -\Theta{\left(1-x-1/\bar{N}\right)} ,
\end{equation}
with $\bar{N}=N e^{\gamma_{\text{E}}}$, $\gamma_{\text{E}}$ the Euler-Mascheroni constant and $\Theta$ the Heaviside step function. Using Eq.~(\ref{eq:large-N integral}), after the change of integration variable $z=1-x$ (which we will often adopt in the following) we obtain
\begin{align}\label{eq:massive resummation-bis}
	\widetilde{\Gamma}^{(4,\text{res})}_{\ell_1=0}(N,\xi)=\left(1+ \frac{\as(q^2)}{\pi}\mathcal{K}_1(\xi)\right) \exp \left[-2\, \gszero(\beta) \int^1_{1/\bar N} \frac{\de z}{z}   \frac{\as\left(z^2 q^2\right)}{\pi}\right].
\end{align}
In this expression, inverse powers of $\bar N$ are systematically neglected, while the dependence on $\frac{m}{q}$ is taken into account exactly. For this reason, 
$\widetilde{\Gamma}^{(4,\text{res})}_{\ell_1}(N,\xi)$ is only accurate for values of $N$ such that $\frac{1}{\bar N}<\frac{m}{q}$. As a consequence, the heavy quark threshold,
corresponding to $z=\frac{m}{q}$, lies within the integration range in the exponent of Eq.~(\ref{eq:massive resummation-bis}). In the variable-flavour number scheme for the running coupling, the number of active flavours that contribute to the QCD $\beta$ function is increased by one unit at each quark mass threshold. Hence,
the running coupling in Eq.~(\ref{eq:massive resummation-bis}) reads
\begin{equation}\label{eq:vfns-coupling}
\as(q^2z^2)=\as^{(n_l)}(q^2z^2)\Theta(\sqrt{\xi}-z)+\as^{(n_f)}(q^2z^2)\Theta(z-\sqrt{\xi}) ,
\end{equation}
where $n_f=n_l+1$.

It is interesting to study the small-$\xi$ behaviour of Eq.~(\ref{eq:gammazero}). We find
\begin{equation} \label{ref:gs0-expanded}
\gszero=-\cf \left( \log \xi+1\right) +\mathcal{O}(\xi).
\end{equation}
Thus, a class of collinear logarithms, specifically those that accompany a soft logarithm, are also resummed by Eq.~(\ref{eq:massive resummation-bis}). However, a full resummation of all mass logarithms can only be performed in a different calculation scheme, as detailed in the next section. 
For later convenience, we introduce a subtracted massive soft anomalous dimension:
\begin{equation}\label{eq:gszero-sub}
	\gszerotilde(\beta)=\gszero+\cf \left( \log \xi+1\right)=
	\cf\left( \frac{1+\beta^2}{2\beta} \log\frac{1+\beta}{1-\beta}+\log\frac{1-\beta^2}{4}\right),
	\end{equation}
which vanishes as $\xi \to 0$, $\beta \to 1$.

Terms proportional to powers of $\log\xi$ with $N$-independent coefficients appear in the prefactor $\mathcal{K}$. In the small-$\xi$ limit we find~\footnote{This coefficient depends on the renormalisation scheme adopted for the Yukawa coupling of the Higgs boson to the $b$-quark. Here we employ the $\msb$ scheme, while in Ref.~\cite{Gaggero:2022hmv} the on-shell scheme was used.}
\begin{equation} \label{eq:Cone}
	\begin{split}
	\mathcal{K}_1(\xi)&= \frac{\cf}{2}\left(\log^2{\xi}-\log{\xi}+\pi^2+\mathcal{O}(\xi)\right).
	\end{split}
\end{equation}
As pointed out in Ref.~\cite{Gaggero:2022hmv}, this is not what one would expect: at order $\as^k$, the leading collinear logarithm should be proportional to $\log^k\xi$.
It was shown in Ref.~\cite{Gaggero:2022hmv} that
the mass appearing in the double logarithm contribution at order $\as$ is the mass of the undetected massive parton, $\bar b$ in the present case. 
This suggests that, conversely, spurious double logarithms of $N$ from radiation off the undetected parton may arise in the 5-flavour scheme calculation.
In the next section, we will confirm this expectation. 

\subsection{Soft resummation in the massless scheme}
\label{sec:massless-resum}
The resummation of soft logarithms for the $b$ energy spectrum in the massless (5-flavour) approach was performed in Ref.~\cite{Cacciari:2001cw} and the specific case of $H \to b \bar b$ was considered in Ref.~\cite{Corcella:2004xv}.
Logarithms of the Mellin variable $N$ appear both in the coefficient functions and in the initial condition for fragmentation functions into heavy quarks, which are typically given at an energy scale of the order of the heavy quark mass. The general expression for the resummed energy spectrum has the form
	\begin{align}\label{fragm-master}
		\widetilde{\Gamma}_{\ell \ell_2}^{(5,\text{res})}(N,\xi)&= \widetilde{\mathcal{C}}\left(N,\frac{\muf^2}{q^2},\frac{\mur^2}{q^2},\as(\mur^2)\right)
		\widetilde{\mathcal{E}}(N,\muOf^2,\muf^2)\, \widetilde{\mathcal{D}}_0\left(N,\frac{\muOf^2}{m^2},\frac{\muOr^2}{m^2},\as(\muOr^2)\right),
	\end{align}
where $\widetilde{\mathcal{C}}$ is the coefficient function, $\widetilde{\mathcal{E}}$ the DGLAP evolution kernel for the fragmentation function, and $ \widetilde{\mathcal{D}}_0$ the initial condition.
For future convenience, it will be useful to separate off the logarithmically enhanced contribution to $\widetilde{\mathcal{E}}$ from the regular contribution at large $N$. To this purpose, we define a subtracted evolution kernel $\widetilde{\mathcal{E}}^{(\text{sub})}(N,\muOf^2,\muf^2)$
through
\begin{equation}
\label{eq:DGLAPsubtracted}
\widetilde{\mathcal{E}}(N,\muOf^2,\muf^2) =
\widetilde{\mathcal{E}}^{(\text{sub})}(N,\muOf^2,\muf^2)  \exp\left[E(N,\muOf^2,\muf^2)\right],
\end{equation}
where
\begin{equation}\label{eq:evolution operator integral}
E(N,\muOf^2,\muf^2)=-\int^{\muf^2}_{\muOf^2} \frac{\de k^2}{k^2}  \left\{A(\as(k^2)) \log \bar{N} + \frac{1}{2}B(\as(k^2)) \right\}.
\end{equation}
By construction, $\widetilde{\mathcal{E}}^{(\text{sub})}\to 1$ as $N\to\infty$.

The renormalisation and factorisation scales $\mur^2,\muf^2$ are chosen to be of the order of magnitude of $q^2$, while the corresponding reference values
 $\muOr^2,\muOf^2$ are of order  $m^2$.

We remind the reader that knowledge of the DGLAP evolution kernel at the $\ell$-loop order allows one to resum collinear logarithms to N$^{\ell}$LL. In this work we consider the case $\ell=1$. 
At NLL in $N$, i.e. \ $\ell_2=1$, the resummed decay rate can be written 
\begin{align}\label{eq:FF-sep}
		\widetilde{\Gamma}_{\ell=1, \ell_2=1}^{(5,\text{res})}(N,\xi)&=
				 \left(1+ \frac{\as(\mur^2) \cf}{\pi}\mathcal{C}_0^{(1)} \right) \left(1+ \frac{\as(\muOr^2) \cf}{\pi}\mathcal{D}_0^{(1)} \right) 
				 \widetilde{\mathcal{E}}^{(\text{sub})}(N,\muOf^2,\muf^2)  \nonumber\\ &
				 \exp \left[J\left(N,\frac{\mur^2}{q^2},\frac{\muf^2}{q^2},\frac{\muOr^2}{m^2},\frac{\muOf^2}{m^2}\right)+\bar{J}\left(N,\frac{\mur^2}{q^2},\frac{\muOr^2}{m^2}\right)\ \right].
\end{align}
The factor $\exp(J)$ in Eq.~(\ref{eq:FF-sep}) describes  soft radiation emitted collinearly to the observed $b$-quark. The jet function $J$ has the following form: 
\begin{equation}\label{eq:j-def}
J=D_0 + E+\Delta,
\end{equation}
where 
\begin{align}\label{eq: delta function}
	\Delta\left(N,\frac{\muf^2}{q^2},\frac{\mur^2}{q^2}\right)=&\int_{\frac{1}{\bar{N}}}^1 \frac{\de z}{z}\int^{\muf^2}_{z^2q^2}\frac{\de k^2}{k^2} A(\as(k^2)),\\
\label{eq: initial condition}
D_0\left(N,\frac{\muOf^2}{m^2},\frac{\muOr^2}{m^2}\right)=& -\int_{\frac{1}{\bar{N}}}^1  \frac{\de z}{z} \left\{ \int^{\muOf^2}_{z^2 m^2} \frac{\de k^2}{k^2} A(\as(k^2))+H\left(\as\left(z^2 m^2\right)\right)\right\}.
\end{align}
Finally, the jet function $\bar{J}$ is related to radiation emitted collinearly to the $\bar{b}$ direction. One finds
\begin{align}
\label{eq: barJ}
	\bar{J}\left(N,\frac{\mur^2}{q^2} \right)=& -\int_{\frac{1}{\bar{N}}}^1  \frac{\de z}{z} \left\{\int^{zq^2}_{z^2q^2}\frac{\de k^2}{k^2} A(\as(k^2))+\frac{1}{2}B(\as\left(zq^2\right))\right\}.
\end{align}
Note that beyond NLL the resummed exponent can no longer be written as the sum of two independent jet functions.

The functions $A,B$ and $H$ have an expansion in powers of $\as$:
 \begin{equation}\label{Resummation_functions}
 	\begin{split}
 		A(\as)=\sum_{k=1}^{\infty} \left(\frac{\as}{\pi}\right)^k A_{k},\quad	B(\as)=\sum_{k=1}^{\infty} \left(\frac{\as}{\pi}\right)^k B_{k},\quad
 		H(\as)=\sum_{k=1}^{\infty} \left(\frac{\as}{\pi}\right)^k H_{k},
 	\end{split}
 \end{equation}
 with
 \begin{equation}\label{Resummation_constants}
 	\begin{split}
 		A_{1}&=\cf,\quad A_{2}^{(n)}=\frac{\cf K^{(n)}}{2}\\
 		B_{1}&=-\frac{3}{2}\cf, \quad H_{1}=-\cf
 	\end{split}
 \end{equation}
 where $K^{(n)}= \ca \left(\frac{67}{18}-\zeta_2\right)-\frac{5}{9} n$. The coefficients given in Eq.~(\ref{Resummation_constants}) are sufficient to achieve NLL accuracy. Explicit expressions for the evolution factor $\mathcal{E}$, the coefficient function $\mathcal{C}$ and the initial condition for the fragmentation function $\mathcal{D}_0$ 
at the relevant level of accuracy are given in App.~\ref{Resummation:FF}.

We now investigate the presence of double logarithmic terms, i.e.\ tems of order $\as^k\log^{k+1}N$, in the exponent of the resummed expression Eq.~(\ref{eq:FF-sep}).
Leading logarithmic contributions to the exponent $J$ originate from the first term in the expansion of the function $A(\as)$ in 
Eqs.~(\ref{eq:evolution operator integral},\ref{eq: delta function},\ref{eq: initial condition}). We find
\begin{align}
J(N,\xi)&=-\frac{A_1}{\pi}\int_{\frac{1}{\bar{N}}}^1\frac{\de z}{z}\left\{\int_{\muOf^2}^{\muf^2}\frac{dk^2}{k^2}\as(k^2)
-\int_{z^2q^2}^{\muf^2}\frac{dk^2}{k^2}\as(k^2)
+\int^{\muOf^2}_{z^2m^2}\frac{dk^2}{k^2}\as(k^2)\right\}+\text{NLL}
\nonumber\\
&=-\frac{A_1}{\pi}\int_{\frac{1}{\bar{N}}}^1\frac{\de z}{z}
\int^{z^2q^2}_{z^2m^2}\frac{dk^2}{k^2}\as(k^2)
+\text{NLL},
\label{JLL}
\end{align}
where we have not indicated the dependence on $\as$ and renormalisation and factorisation scales to simplify notations.
Let us first evaluate the expression above at fixed coupling. We find
\begin{equation}\label{JNLL-fc}
J(N,\xi)=-\frac{A_1\as}{\pi} \log \bar N \, \log \xi +\order{\as^2}
\end{equation}
which shows that here are no double logarithms of $N$ at this level of accuracy. An analogous calculation for the jet function related to the unmeasured $\bar b$ leads to 
\begin{equation}\label{eq:Jbar-Nspace}
\bar J(N)=-\frac{A_1}{\pi}\int_{\frac{1}{\bar N}}^1\frac{\de z}{z}\int^{zq^2}_{z^2q^2}\frac{\de k^2}{k^2} \as(k^2)+\text{NLL}
=-\frac{A_1\as}{2\pi}\log^2\bar N+\order{\as^2}.
\end{equation}
In this case double logarithms of $N$ appear in the $\bar J$ contribution to the exponent already at order $\as$.
We note that, while the calculation of $D_0$, which contributes to the measured jet function $J$, is performed in the so-called quasi-collinear limit~\cite{Catani:2000ef,Catani:2002hc}, the calculation of $\bar J$ is performed in the pure collinear limit. We will expand on this consideration in the following section, showing that the treatment of $J$ and $\bar J$ on equal footing will allow us to consistently match $\widetilde{\Gamma}^{(4,\text{res})}_{\ell_1=0}$ and $\widetilde{\Gamma}_{\ell=1, \ell_2=1}^{(5,\text{res})}$ to all orders.

\begin{figure}
	\centering
	\includegraphics[width=0.4\textwidth,page=1]{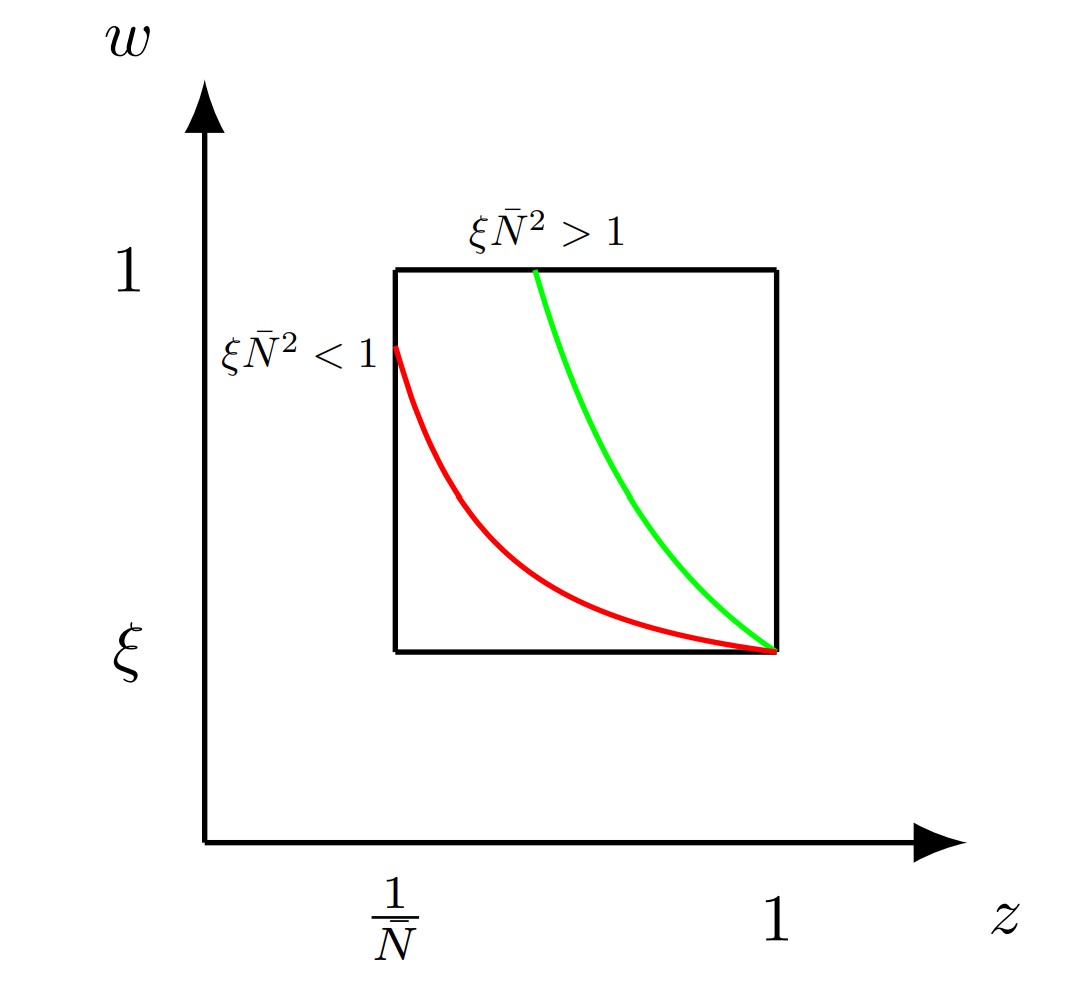}
	\caption{The integration region in Eq.~(\ref{eq:JNLL2}). The threshold curve $w=\frac{\xi}{z^2}$ is also shown, for $\xi\bar N^2<1$ (red curve) and $\xi\bar N^2>1$ (green curve).}
	\label{intdom}
\end{figure}

The logarithmic structure of Eq.~(\ref{JLL}) is slightly modified when the running of the strong coupling is taken into account.
The integration region in Eq.~(\ref{JLL}) is mapped into a rectangle by  the integration variable change
$w=\frac{k^2}{z^2 q^2}$:~\footnote{Note that the variable $w$ is related to the rapidity of the emission with respect to the direction of the hard momentum. We will come back to this observation in Sect.~\ref{sec:jet_mom}, where we perform the calculation in momentum space.}
\begin{equation}
J(N,\xi)=-\frac{A_1}{\pi}\int_{\frac{1}{\bar{N}}}^1\frac{\de z}{z}
\int_\xi^1\frac{dw}{w}\as(q^2 z^2 w)
+\text{NLL}.
\label{eq:JNLL2}
\end{equation}
The integration domain is divided in two regions by the curve $q^2 z^2 w=m^2$, or $w=\frac{\xi}{z^2}$; in the lower region
$\as=\as^{(4)}$, while in the upper region $\as=\as^{(5)}$ (see Fig.~\ref{intdom}). 
We must distinguish two cases. 
If $\xi\bar N^2>1$ (green curve in Fig.~\ref{intdom}), then
\begin{align} 
J(N,\xi)_{\xi\bar N^2>1}&=-\frac{A_1}{\pi}
\int_{\frac{1}{\bar{N}}}^{\sqrt{\xi}}\frac{\de z}{z}
\int_\xi^1\frac{dw}{w}\as^{(4)}(q^2 z^2 w)
\nonumber\\
&-\frac{A_1}{\pi}\int_{\sqrt{\xi}}^1\frac{dz}{z}
\left[\int_\xi^{\frac{\xi}{z^2}}\frac{dw}{w}\as^{(4)}(q^2 z^2 w)
+\int_{\frac{\xi}{z^2}}^1\frac{dw}{w}\as^{(5)}(q^2 z^2 w)\right]
+\text{NLL}.
\label{eq: J without LL}
\end{align}
The terms in the last line are $N$-independent. 
Using the leading logarithmic solution
for $\as^{(n)}(\mu^2)$ 
\begin{equation}
\as^{(4)}(q^2 z^2 w)=\sum_{k=1}^\infty c_k\left[\as^{(4)}(q^2 w)\right]^k\log^{k-1}z;\qquad
c_k=\left[-2\beta_0^{(4)}\right]^{k-1}
\end{equation}
we get
\begin{equation}
J(N,\xi)_{\xi\bar N^2>1}=\frac{A_1}{\pi}\sum_{k=1}^\infty \frac{c_k}{k}
\int_\xi^1\frac{dw}{w}\left[\as^{(4)}(q^2 w)\log\frac{1}{\bar N}\right]^k+\text{NLL},
\end{equation}
which does not contain any leading logarithmic term, i.e.\ tems of order $\as^k\log^{k+1}N$.

On the other hand, if $\xi\bar N^2<1$ (red curve in Fig.~\ref{intdom}), then
\begin{equation}\label{eq:Jnspace_region1}
J(N,\xi)_{\xi\bar N^2<1}=-\frac{A_1}{\pi}\int_{\frac{1}{\bar{N}}}^1\frac{\de z}{z}\left[
\int_\xi^{\frac{\xi}{z^2}}\frac{dw}{w}\as^{(4)}(q^2 z^2 w)
+\int_{\frac{\xi}{z^2}}^1\frac{dw}{w}\as^{(5)}(q^2 z^2 w)\right]
+\text{NLL}.
\end{equation}
In this case, terms of order $\as^k\log^{k+1}N$ cancel at order $\as$, but not at higher orders. Indeed, using the leading logarithmic solution
for $\as^{(n)}(\mu^2)$ with the threshold condition
\begin{equation}
\as^{(n)}(m^2)=\as^{(n+1)}(m^2)\equiv\as(m^2),
\end{equation}
and expanding in powers of $\as(m^2)$ one finds
\begin{align}
&J(N,\xi)_{\xi\bar N^2<1}
\nonumber\\
&=\frac{A_1}{\pi}\int_{\frac{1}{\bar{N}}}^1\frac{\de z}{z}\left[
\frac{1}{\beta_0^{(4)}}\log\left(1+\as(m^2)\beta_0^{(4)}\log z^2\right)
-\frac{1}{\beta_0^{(5)}}\log\left(1+\as(m^2)\beta_0^{(5)}\log\frac{z^2}{\xi}\right)
\right]
+\text{NLL}
\nonumber\\
&=\frac{A_1}{\pi}\left[\as(m^2)\log\bar{N}\log\xi
+\frac{2}{3}
\left(\beta_0^{(4)}-\beta_0^{(5)}\right)\as^2(m^2)\log^3\bar{N}
+\mathcal{O}(\as^3\log^4\bar{N})\right]
+\text{NLL}.
\end{align}
Thus, we have found that the jet function $J$ also exhibits leading logarithms of $N$, because of running coupling effects, but only in the region $\bar N <1/\sqrt{\xi}$. 

We note that the above results refer to the case of real and positive Mellin moments. When the resummed spectrum in Mellin space is continued analytically to the complex $N$ plane, which is mandatory for Mellin inversion, their interpretation is not straightforward, especially as far as the hierarchy between different energy scales is concerned.
A better understanding of this issue can be obtained by looking first at resummation in momentum space, as we show in the next section.

\section{Jet functions in momentum space} \label{sec:jet_mom}

In this section, the calculations presented above are performed directly in momentum space. This calculation will allow us to obtain a clearer picture of the different scales at play and, hence, of the different kinematic regions we have to consider. 
Armed with such an understanding, we will be able to obtain a Mellin-space resummation formula that consistently resums both logarithms of $N$ and logarithms of $\xi$ to NLL accuracy, in all relevant phase-space regions. 

We will also discuss a Lund plane representation of the kinematics. 
Lund diagrams~\cite{Andersson:1988gp} are a useful way to represent the available phase space for the emission of soft and/or collinear gluons off a hard dipole. 
This approach has been proven particularly useful for hadronic final-states resummation, in the presence of multiple scales: it has been successfully applied to the NLL resummation of event shapes~\cite{Banfi:2004yd}, and, more recently, of a variety of jet observables, see e.g.~\cite{Marzani:2019hun}.

\subsection{Resummation in momentum space}
Let us briefly summarise how to perform NLL resummation in momentum space.~\cite{Banfi:2004yd} In this work we are interested in the energy spectrum of a final-state heavy quark, but the technique we are illustrating applies to a more general class of observables. We therefore consider 
of a generic infrared and collinear (IRC) safe observable $\mathcal{V}$, a positive function of final-state momenta that vanishes at Born level.
It proves convenient to consider the normalised cumulative  distribution
\begin{equation}
\Sigma(v) = \frac{1}{\Gamma_0}\int_0^v \de v' \frac{\de \Gamma }{\de v'},
\end{equation}
i.e.\ the probability for the observable $\mathcal{V}$ to be smaller than some given value $v$, rather than the differential rate $\frac{\de\Gamma}{\de v}$.
Because we are ultimately interested in $H \to b \bar b$, we focus on a single quark-antiquark hard dipole with centre-of-mass momenta $p_1$ and $p_2$ respectively. We first address the case of massless quarks, $p_1^2=p_2^2=0$, and then extend our results to massive quarks.

We begin by considering the $\mathcal{O}(\as)$ contribution to $\Sigma(v)$, corresponding to one-gluon emission (plus $\mathcal{O}(\as)$ virtual corrections.)  We denote by $k$ the momentum of the emitted gluon.
Neglecting for the moment the contribution of virtual corrections, the normalised distribution at order $\as$ can be schematically written
\begin{equation}
\frac{1}{\Gamma_0}\frac{\de\Gamma}{\de v}=\delta(v)+\int \de\Gamma^{(1)}\,\delta\left(\mathcal{V}(p_1,p_2,k)-v\right),
\end{equation}
where $d\Gamma^{(1)}$ is the differential width for one-gluon emission.
Therefore
\begin{align}
\Sigma(v)&=1+\int \de\Gamma^{(1)}\,\int_0^v \de v'\,\delta\left(\mathcal{V}(p_1,p_2,k)-v'\right)
\nonumber\\
&=1+\int \de\Gamma^{(1)}\,\Theta\left(v-\mathcal{V}(p_1,p_2,k)\right).
\end{align}

In the massless case, QCD emission probabilities, in the soft and collinear limits, behave as $\frac{\de k_{t}}{k_{t}} \de \eta_i$, where $k_t$ and $\eta_i$ are the transverse momentum 
and rapidity (assumed positive) of the emission with respect to emitting particle momentum direction:
\begin{equation}
k=(k_t\cosh\eta_i,\vec k_t,k_t\sinh\eta_i).
\label{eq:gluonmom}
\end{equation}
It is therefore convenient to express the kinematics in terms of these variables. 
The boundary of the phase space in terms of $k_t,\eta_i$ is determined
by the condition
\begin{equation}
(q-k)^2=q^2-2qk_t\cosh\eta_i\ge 0
\label{psbounds}
\end{equation}
which gives
\begin{equation}
0\le k_t\le\frac{q}{2};\qquad
0\le\eta_i\le\eta_{\rm max};\qquad
\eta_{\rm max}=\log\left[\frac{q}{2k_t}\left(1+\sqrt{1-\frac{4k_t^2}{q^2}}\right)\right].
\label{psbounds2}
\end{equation}
In our case, logarithmically enhanced contributions to the cumulative distribution  originate from gluon emission
either collinear to $p_1$ or to $p_2$. We therefore define two regions of collinear emission:
\begin{equation}
\eta_{\rm cut}<\eta_i<\eta_{\rm max};\qquad i=1,2
\label{etacoll}
\end{equation}
in terms of some large rapidity value $\eta_{\rm cut}$.
The condition (\ref{etacoll}) implies
\begin{equation}
0<k_t<\frac{q}{2\cosh\eta_{\rm cut}}
\end{equation}
as a consequence of Eq.~(\ref{psbounds}). For large values of $\eta_{\rm cut}$ we have
\begin{equation}
0<k_t<qe^{-\eta_{\rm cut}};\qquad \eta_{\rm max}=-\log\frac{k_t}{q}.
\end{equation}
Restricting the integration region to the collinear-emission regions of the phase space, we have 
\begin{align}
\Sigma(v)=1&+\int_0^{Q^2}\de k_t^2\,\int_{-\log\frac{Q}{q}}^{-\log\frac{k_t}{q}}\de\eta_1\,|\mathcal{M}|^2\Theta\left(v-\mathcal{V}(p_1,p_2,k)\right)
\nonumber\\
&+\int_0^{Q^2}\de k_t^2\,\int^{-\log\frac{k_t}{q}}_{-\log\frac{Q}{q}}\de\eta_2\,|\mathcal{M}|^2\Theta\left(v-\mathcal{V}(p_1,p_2,k)\right)
\label{cumulative}
\end{align}
where $\mathcal{M}$ is the appropriate invariant amplitude, and $Q$ is the maximum value allowed for the transverse momentum,
\begin{equation}
Q=qe^{-\eta_{\rm cut}}.
\end{equation}

It is shown in Ref.~\cite{Banfi:2004yd} that in most cases the observables of interest  behave, in the soft and collinear limits, as
\begin{equation}
\mathcal{V}(p_1,p_2,k)\to d_i\left(\frac{k_t}{Q}\right)^{a_i} e^{-b_i\eta_i},
\label{Vcoll}
\end{equation}
where we have chosen $Q$ as reference (hard) scale, and the index $i$ labels the different collinear regions (in our case, the two regions collinear to either $\vec p_1$ or $\vec p_2$.)
The positive constants $a_i,b_i,d_i$ depend in general on the particular region of the phase space where the collinear limit is taken; in our case, we have two sets of constants. 
In the region collinear to the quark momentum $p_1$ we have, in the large rapidity limit,
\begin{equation}
\mathcal{V}=V_1(k_t,\eta_1)=d_1 \,\left(\frac{k_t}{Q}\right)^{a_1} e^{-b_1\eta_1}.
\end{equation}
Similarly, in the region collinear to the antiquark momentum 
\begin{equation}
\mathcal{V}= V_2(k_t,\eta_2)=d_2 \,\left(\frac{k_t}{Q}\right)^{a_2} e^{- b_2\eta_2}
\end{equation}

The squared amplitudes that appear in Eq.~(\ref{cumulative}) can be computed,  in the collinear limits, by exploiting factorization of collinear singularities.
Let us first consider the region where the gluon is emitted in the direction of the quark momentum. It is convenient to adopt a Sudakov parametrisation of the gluon momentum:
\begin{equation}\label{eq:sudakov1}
k=z_1 p_1+\bar z_1\bar p_1+\kappa,
\end{equation}
where $\bar p_1=(p_1^0,-\vec p_1)$ and $\kappa$ is a spacelike vector such that $p_1 \cdot \kappa=\bar p_1 \cdot \kappa=0$, and $\kappa^2=-k_t^2$. The coefficient $\bar z_1$ is suppressed in the collinear limit. Indeed, the gluon mass-shell condition gives
\begin{equation}
k^2=2 z_1 \bar z_1 p_1\cdot\bar p_1-k_t^2=0.
\label{massshell}
\end{equation}
As a consequence, 
\begin{equation}
\bar z_1=\frac{k_t^2}{2p_1\cdot \bar p_1  z_1}=\frac{k_t^2}{q^2 z_1}+\mathcal{O}\left(\frac{k_t^3}{q^3}\right)
\end{equation}
is much smaller than $z_1$ in the small-$k_t$ limit. In this limit, $z_1$ represents the fraction of the quark energy which is carried away by the gluon. The gluon rapidity with respect to the quark momentum direction is given by
\begin{equation}
\eta_1= \frac{1}{2}\log \frac{z_1}{\bar z_1}\simeq\log z_1-\log\frac{k_t}{q}.
\label{etaplus}
\end{equation}
Conversely, when the emission is collinear to the antiquark, we may write the gluon momentum as
\begin{equation}
k= z_2 p_2 + \bar z_2 \bar p_2 + \kappa,
\label{eq:sudakov2}
\end{equation}
with $\bar p_2=(p_2^0,-\vec p_2)$. We get
\begin{equation}
\bar z_2=\frac{k_t^2}{2p_2\cdot \bar p_2  z_2}=\frac{k_t^2}{q^2 z_2}+\mathcal{O}\left(\frac{k_t^3}{q^3}\right),
\end{equation}
and 
\begin{equation}
\eta_2=\frac{1}{2}\log\frac{z_2}{\bar z_2}\simeq\log z_2-\log\frac{k_t}{q}.
\label{etaminus}
\end{equation}

Thus, we have
\begin{align}
\Sigma(v)= 1&+
\sum_{i=1,2}
\int_0^{Q^2}\frac{\de k_t^2}{k_t^2} \int_{\frac{k_t}{Q}}^1  \de z_i\, \frac{\as(k_t^2)}{2\pi} P_{qq} (1-z_i) 
\left[ \Theta\left( v-  V_i(k_t,\eta_i)\right) -1\right],
\label{eq:one-emission-massless-start}
\end{align}
where 
\begin{equation}\label{eq:masslessAP}
	P_{qq}(1-z_i)= \cf \frac{1+(1-z_i)^2}{z}=P_{gq}(z_i)
\end{equation}
is the appropriate timelike DGLAP splitting function. The $-1$ contributions inside square brackets in Eq.~(\ref{eq:one-emission-massless-start}) account for virtual corrections, and IRC safety of the observable ensures the cancellation of the $z_i \to 0$ and $k_{t} \to 0$ singularities. 
Eq.~(\ref{eq:one-emission-massless-start}) can therefore be written 
\begin{align}
\Sigma(v)= 1&-\sum_{i=1,2}\int_0^{Q^2}\frac{\de k_t^2}{k_t^2} \int_{\frac{k_t}{Q}}^1  \de  z_i\, \frac{\as(k_t^2)}{2\pi} P_{gq} (z_i) 
\Theta\left(V_i(k_t,\eta_i)-v\right).
\label{eq:one-emission-massless}
\end{align}
The choice of the reference hard scale of the process $Q$, and therefore of $\eta_{\rm cut}$, is immaterial at NLL~\cite{Banfi:2004yd}. 
Following the literature we choose $\eta_{\rm cut}=0$. Note that with this choice the upper bound for $k_t$ is larger than its kinematic limit $q/2$. However, this phase-space region is outside the jurisdiction of the resummed calculation and proper energy-momentum conservation is usually restored when matching to a fixed-order calculation.

At LL accuracy, where each emission comes with a maximal number of
logarithms, one can further assume that a single emission strongly dominates the value of the observable, leading to the exponentiation of the one-loop calculation also in momentum space (see for instance~\cite{Marzani:2019hun}):
\begin{align}\label{eq:LL}
\Sigma_\text{LL}(v) =\exp \left[\mathcal{R} (v) \right],
\end{align}
where the Sudakov form factor $\mathcal{R} (v)$, which represents the no-emission probability, is expressed as the sum of two jet functions, one for each collinear sector:
\begin{align}
 \label{eq: Sudakov}
\mathcal{R} (v)&= j(v)+\bar j(v),
\end{align}
with
\begin{align}
j(v)&=- \int_0^1 \de z_1\int_0^{q^2} \frac{\de k_t^2}{k_t^2} \frac{\as(k_t^2)}{2\pi} P_{gq} (z_1)\Theta(\eta_1)  \Theta\left(  V_1 (k_t,\eta_1)-v \right)
\nonumber\\
 \bar j(v)&=- \int_0^1 \de \bar z_2 \int_0^{q^2} \frac{\de k_t^2}{k_t^2} \frac{\as(k_t^2)}{2\pi} P_{gq} ( z_2)\Theta(\eta_2)\Theta\left(  V_2(k_t,\eta_2)-v \right).
\label{eq: Sudakov-2}
\end{align}

In order to achieve full NLL accuracy, Eq.~(\ref{eq:LL}) must be corrected in two ways. First, one should consider running of the strong coupling in Eq.~(\ref{eq: Sudakov-2}) at two loops in the so-called Catani-Marchesini-Webber (CMW) scheme:~\cite{Catani:1990rr}
\begin{equation} \label{eq:as_CMW}
\as(k_t^2) \to \as^{\text{CMW}}(k_t^2)=\as(k_t^2)\left(1+\as(k_t^2) \frac{K^{(n)}}{2\pi}\right),
\end{equation}
where $\as(k_t^2)$ is in the decoupling scheme, see Eq.~(\ref{eq:vfns-coupling}). 
Second, we can no longer work in the strongly-ordered approximation, and the resummation must be performed either with numerical methods~\cite{Banfi:2004yd} or  in a conjugate (e.g.\ Mellin) space, in order to factorise the observable definition. In the latter case, at the end of the calculation, the result should then be brought back to physical space. In some cases, this inversion can be done, to a given logarithmic accuracy, in closed-form, resulting in a correction factor, which only depends on derivatives of the Sudakov form factor.~\cite{Catani:1992ua} However, we note that Eq.~(\ref{eq:LL}), supplemented by the prescription in Eq.~(\ref{eq:as_CMW}), is enough to achieve NLL accuracy in Mellin space, provided that we identify $v=\bar N^{-1}$, as shown in App.~\ref{sec:moments}. So finally
\begin{align}\label{eq:NLL}
\widetilde{\Sigma}_\text{NLL}(N) =\exp \left[j(v)+ \bar j(v)\right]\Big|_{v=\bar N^{-1}},
\end{align}
where the replacement (\ref{eq:as_CMW}) in the jet functions $j$ and $\bar j$ is understood.

The formalism described above can be generalised to the case of massive quarks. When $m\ne 0$, kinematics undergoes some minor modifications. The boundaries of the phase space gets slightly modified: the inequality (\ref{psbounds}) becomes
\begin{equation}
(q-k)^2=q^2-2qk_t\cosh\eta_i\ge 4m^2
\label{psboundsmass}
\end{equation}
which gives
\begin{align}
&0\le k_t\le\frac{q}{2}(1-4\xi);\qquad
0\le\eta_i\le\eta_{\rm max}
\nonumber\\
&\eta_{\rm max}=\log\left[\frac{q}{2k_t}\left(1-4\xi+\sqrt{(1-4\xi)^2-\frac{4k_t^2}{q^2}}\right)\right].
\label{psbounds2mass}
\end{align}
Furthermore, the relationship between $z_i$ and $\bar z_i$, which originates from the mass-shell condition for the gluon momentum,
is now given by
\begin{equation}
\bar z_i=-z_i\frac{p_i\cdot \bar p_i}{m^2}\left[1-\sqrt{1+\frac{k_t^2m^2-m^4 z_i^2}{z_i^2(p_2\cdot\bar p_2)^2}}\right]
=\frac{k_t^2}{q^2z_i}+O\left(\xi^2,\frac{k_t^4}{q^4}\right).
\end{equation}
Finally, the expressions Eqs.~(\ref{etaplus},{\ref{etaminus}) for the gluon rapidity in the collinear regions still hold, up to corrections of higher order in $\frac{m}{q}$ and $\frac{k_t}{q}$.

It was realised long ago that squared QCD matrix elements with massive partons factorise in the so-called quasi-collinear limit,~\cite{Catani:2000ef,Catani:2002hc} in which both $\frac{k_t^2}{q^2}$ and $\frac{m^2}{q^2}$ go to zero, with their ratio kept constant. In this limit, the squared invariant amplitude for one-gluon emission takes the form
\begin{equation}\label{eq:quasi-coll-limit}
	|\mathcal{M}|^2\simeq 8\pi\as \frac{z_i(1-z_i)}{k_{t}^2+z_i^2 m^2} P_{bb}\left(1-z_i,k_{t}^2\right)
\end{equation}
with
\begin{equation}\label{eq:Massive AP}
	P_{bb}(1-z_i,k_t^2)= \cf\left(\frac{1+(1-z_i)^2}{z_i}-\frac{2 z_i(1-z_i)m^2}{k_{t}^2+z_i^2 m^2}\right)=P_{gb}(z_i,k_t^2).
\end{equation}
The mass-dependent shift in the denominator of Eq.~(\ref{eq:quasi-coll-limit}) acts as an effective lower bound of a logarithmic $k_t$ integration:
\begin{equation}\label{eq:mass boundary}
\int_0^{q^2}\frac{\de k_{t}^2}{k_{t}^2+z_i^2 m^2}P_{gb}(z_i,k_t^2) =\int_{m^2 z_i^2}^{q^2}\frac{\de k_{t}^2}{k_{t}^2}P_{gb}(z_i,k_t^2-m^2z_i^2)+ \mathcal{O}(\xi).
\end{equation}
Using Eqs.~(\ref{etaplus}) and~(\ref{etaminus}), we obtain $\eta_i<- \frac{1}{2} \log \xi$.
This is the well-known dead-cone effect,~\cite{Dokshitzer:1991fd,Dokshitzer:1995ev} i.e.\ the fact that radiation off massive partons at angles below $m/q$ is not logarithmically enhanced.
Finally, we note that now the observable $\mathcal{V}$ can now explicitly depend on the heavy-flavour mass through $\xi$.

 \subsection{Lund plane interpretation}
The procedure illustrated in the previous subsection is conveniently interpreted in terms of Lund diagrams. 
Lund diagrams offer a graphical representation of the kinematics described above. The Lund plane is usually represented in terms of pairs of the aforementioned logarithmic variables, so that, in the soft and collinear limit, equal areas correspond to equal emission probabilities.
As an example, we show in Fig.~\ref{fig:LundPlane-mass-gen}, on the left, the Lund plane in case of a massless dipole in the dipole rest frame. 

The variable in the vertical axis is the logarithm of $\frac{k_t^2}{q^2}$. The variable on the horizontal axis in the right half-plane is the rapidity $\eta_2$ of the emission with respect to the antiquark momentum, while on the left we report the rapidity $\eta_1$ with respect to the quark momentum.
Note that
\begin{equation}\label{eq:eta-angle}
\eta_i=\frac{1}{2}\log\frac{\sqrt{1+\frac{k_t^2}{k_\ell^2}}+1}{\sqrt{1+\frac{k_t^2}{k_\ell^2}}-1}
=\frac{1}{2}\log\frac{1+\cos\theta_i}{1-\cos\theta_i},
\end{equation}
where $k_\ell$ is the longitudinal component of the gluon momentum along the direction of parton $i$, and $\theta_i$ the angle between the gluon and the parton momenta. Hence, in the collinear limits
$\cos\theta_i\to 1$
\begin{align}
\eta_i\to -\log\frac{\theta_i}{2}.
\end{align}
This is explicitly indicated on the horizontal axis in Fig.~\ref{fig:LundPlane-mass-gen}.
In the massless case the boundaries of the phase space, namely the lines $z_1=1$ and $z_2=1$, are represented on the Lund plane by the lines $\log\frac{k_t^2}{q^2}=2\eta_i$.
Lines of constant $\mathcal{V}$ in the soft and collinear limits, when Eq.~(\ref{Vcoll}) applies, are represented on the Lund plane by straight lines, as shown in Fig.~\ref{fig:LundPlane-mass-gen}.

\begin{figure}[htb]
	\centering
\includegraphics[width=0.49\textwidth,page=1]{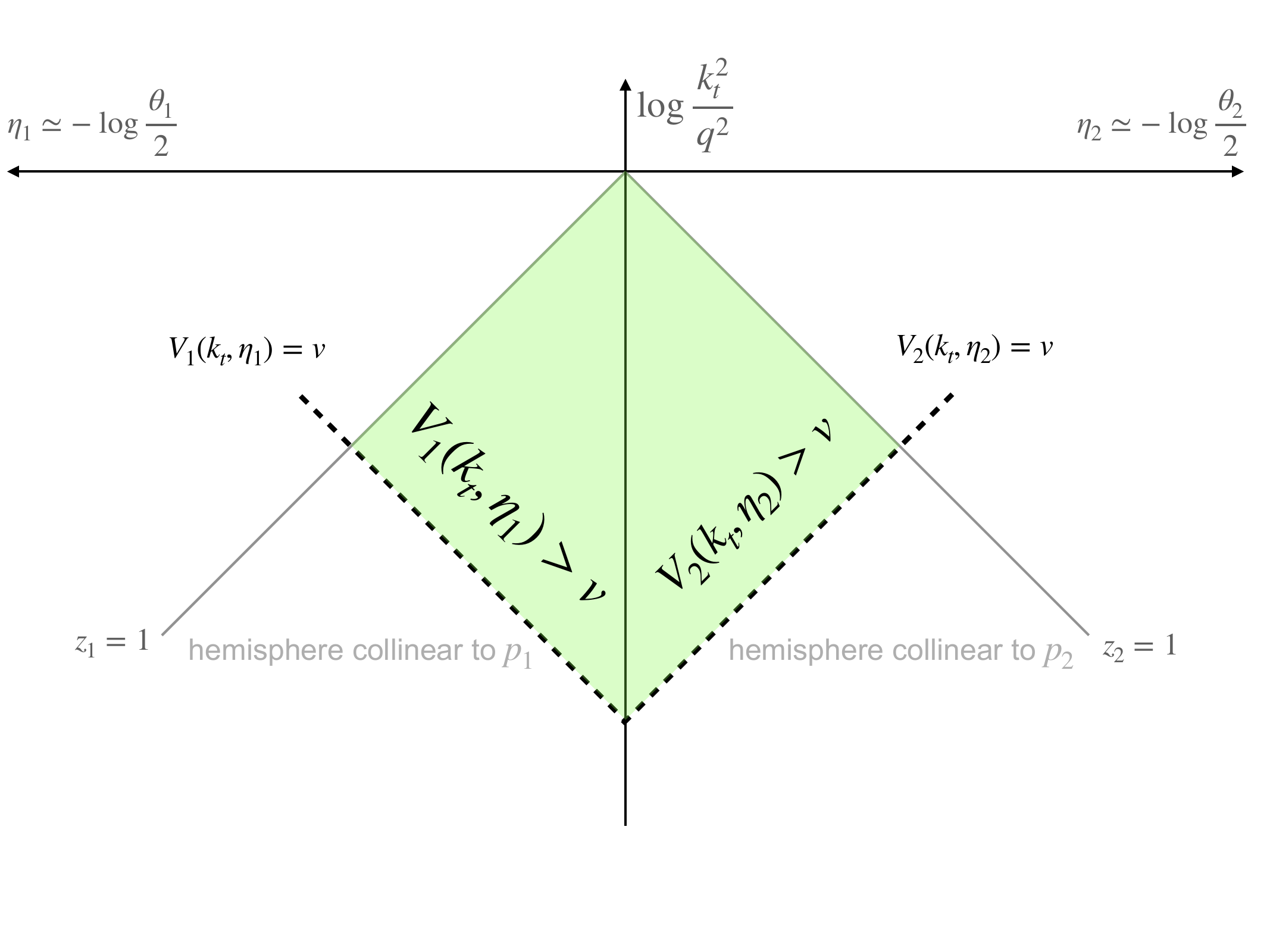}
\includegraphics[width=0.49\textwidth,page=2]{figures/lund.pdf}
	\caption{Lund plane representation of a soft-collinear emission off a hard quark-antiquark dipole, in the dipole rest frame for massless (left) and massive (right) quarks.
	The observable $\mathcal{V}$ under study takes a constant value $v$ along
solid dashed lines. The shaded area represents the region where the cumulative distribution differs from zero.
The vertical lines $\eta_i= -\frac{1}{2} \log \xi$ indicate the position of the dead cone. 
The line $k_t^2=\xi q^2=m^2$ marks the separation between regions of phase-space with $4$ or $5$ active flavours.
	}
	\label{fig:LundPlane-mass-gen}
\end{figure}

The cumulative distribution also has a graphic interpretation on the Lund plane.
The real emission cannot take place everywhere because it must give a contribution to the observable smaller than $v$. Thus, the emission probability corresponding to the phase-space region above the lines $V_1=v$ and $V_2=v$, which is indicated with the shaded area in Fig.~\ref{fig:LundPlane-mass-gen}, is subtracted from 1 in Eq.~(\ref{eq:one-emission-massless}).
Thus, in order to compute the Sudakov form factor $\mathcal{R}$, which represent the no-emission probability, we have to integrate the splitting functions (with running coupling at the appropriate accuracy, as previously discussed) over the shaded areas in Fig.~\ref{fig:LundPlane-mass-gen}.

In the massive case, we have to employ the quasi-collinear limits of matrix elements squared, Eq.~(\ref{eq:quasi-coll-limit}). This would seem to spoil the Lund plane interpretation. 
However, as noticed in Eq.~(\ref{eq:mass boundary}), from the point of view of the logarithmic structure, we can still use the massless phase-space (but with quasi-collinear splitting functions) and the quark mass acts as a new phase-space boundary. 
On the Lund plane, this allowed region is bounded by the two vertical lines at $\eta_i=-\frac{1}{2} \log \xi$. For rapidities between these values, the heavy quark mass can be neglected altogether, while in the collinear regions $\eta_i>-\frac{1}{2} \log \xi$ the quark mass acts as a cut-off for collinear singularities. This is shown in the right panel of 
Fig.~\ref{fig:LundPlane-mass-gen}.
It is also useful to mark on the Lund plane the boundary between the regions of phase-space with $4$ or $5$ active flavours. This transition takes place when $k_t^2 =m^2= \xi q^2$.

We have already noticed that, because the observable $\mathcal{V}$ can depend explicitly on the heavy-quark mass, we do not expect $V_1(k_t,\eta_1,\xi)$ and $V_2(k_t, \eta_2, \xi)$ to be of the same form as Eq.~(\ref{Vcoll}). Consequently, on the Lund plane, lines of constant $1-x$ are deformed with respect to the massless case. However,  we are still allowed to represent them as solid-dashed straight lines on the Lund plane, provided that we take into account the boundaries at $\eta_i =-\frac{1}{2} \log \xi$, as shown in the right panel of Fig.~\ref{fig:LundPlane-mass-gen}.

\subsection{Resummation of the cumulative distribution}
In this section, we specialise the above discussion to the case of the observable $\mathcal{V}=1-x$, with $x$ defined in Eq.~(\ref{eq:xdef}).
In order obtain a resummed result for the cumulative distribution $\Sigma(1-x)$, we must discuss the parametrisations $V_i$, $i=1,2$ of the observable $\mathcal{V}=1-x$.
Using energy-momentum conservation, we find
\begin{equation} \label{eq:1massless-start}
1-x= \frac{2 p_2 \cdot k}{q^2},
\end{equation}
which means that a measurement of $1-x$ is equivalent to a measurement of the invariant mass of the $\bar b  g$ system recoiling against the $b$ quark. 
The Sudakov parametrisations (\ref{eq:sudakov1},\ref{eq:sudakov2}) give
\begin{align}
1-x&=\frac{2p_2\cdot p_1 z_1+2p_2\cdot\bar p_1\bar z_1}{q^2}
=z_1+\mathcal{O}\left(\frac{k_t}{q},\frac{m}{q}\right) \quad\text{emission collinear to the quark},
\\
1-x&=\frac{2p_2\cdot p_2 z_2+2p_2\cdot\bar p_2\bar z_2}{q^2}=\xi z_2+\frac{k_t^2}{q^2 z_2}
+\mathcal{O}\left(\frac{k_t^3}{q^3},\frac{k_t^2m}{q^3},\frac{k_t m^2}{q^3},\frac{m^3}{q^3}\right) \\  &\; \;\quad\quad\quad\quad\quad\quad\quad\quad\quad\quad\quad\quad\quad\quad\quad\quad\quad\quad\quad\quad\text{emission collinear to the antiquark}. \nonumber
\end{align}
As we have anticipated, because of the presence of the mass term $\xi$,  the above parametrisation has not the same form as Eq.~(\ref{Vcoll}). 
Consequently, on the Lund plane, lines of constant $1-x$ would be deformed with respect to the massless case. However, to the accuracy we are working at, we are allowed to shift the transverse momentum as described in Eq.~(\ref{eq:mass boundary}), provided that we take into account the boundaries imposed by the dead-cone. 
Thus, to NLL, we can use the following parametrisation:
\begin{equation} \label{eq:obs-massive-final}
1-x=
\begin{cases}
V_1( k_t, \eta_1)=z_1=\frac{k_t}{q}e^{\eta_1}, \\
V_2(k_t,\eta_2)=\frac{k_t^2}{q^2 z_2}=\frac{k_t}{q} e^{-\eta_2},
\end{cases} 
\end{equation}
where we have used Eqs.~(\ref{etaplus}) and~(\ref{etaminus}). Note that this parametrisation coincides with the one one would find in the massless case. Therefore, to the accuracy we are working at, we can represent lines of constant $1-x$ on the Lund plane as straight lines, parallel to the line $z_1=1$.

In order to obtain the resummed cumulative distribution, we compute the Sudakov form factor Eq.~(\ref{eq: Sudakov}). 
In particular, we have to evaluate the measured jet function $j$ and the unmeasured (or recoil) jet function $\bar j$ in momentum space. Using the shift Eq.~(\ref{eq:mass boundary}) and the parametrisation~(\ref{eq:obs-massive-final}), we have
\begin{align}\label{eq:jet-function-J}
j(1-x,\xi)&= -\int_0^1 \de z_1 \int_{z_1^2 m^2}^{q^2} \frac{\de k_t^2}{k_t^2} \frac{\as^\text{CMW}(k_t^2)}{2\pi} P_{gb} (z_1,k_t^2-z_1^2 m^2) \Theta\left(\eta_1\right)   \Theta\left(z_1 -(1-x) \right), \\ \label{eq:jet-function-Jbar}
\bar j(1-x,\xi) &=-\int_0^1 \de z_2\int_{z_2^2 m^2}^{q^2} \frac{\de k_t^2}{k_t^2} \frac{\as^\text{CMW}(k_t^2)}{2\pi} P_{gb} (z_2,k_t^2-z_2^2 m^2)\Theta\left( \eta_2\right)  \Theta\left( \frac{k_t^2}{q^2 z_2} -(1-x)\right).
\end{align}
Note that, in principle, we should also shift the argument of the running coupling. However, as shown in App.~\ref{app:running_coupling}, this effect only contributes beyond NLL accuracy. 

In the following sections, we will express the jet functions as integrals over the running coupling. When this integrals are computed explicitly, both jets functions acquire an explicit dependence on the renormalisation scales $\muOr$, $\mur$, for the 4- and 5-flavour components, respectively. In addition, the measured jet function also depends on the factorisation scales $\muOf$, $\muf$. This dependence is omitted in Sects.~\ref{sec:j-lund} and~\ref{sec:jbar-lund}, but reinstated when we present our final results in App.~\ref{app:calculations}.

As a final remark, we note that, strictly speaking,  the observable $1-x$ is not IRC safe, because emissions that are arbitrarily collinear to the quark carry away some energy. However, because we are considering emissions off a massive quark, collinear singularities do not appear, and are replaced by large logarithmic mass corrections. With the formalism discussed in this section, we can only resum mass logarithms that have a non-vanishing coefficient when $x \to 1$;  the  resummation of mass logarithms is achieved by DGLAP evolution, discussed in Sect.~\ref{sec:massless-resum}.

\subsubsection{Calculation of the measured jet function}\label{sec:j-lund}
In this section we compute the measured jet function in momentum space $j(v,\xi)$ for $v=1-x$, and  we compare it to the analogous quantity $J(N,\xi)$ 
computed in Mellin space in Sect.~\ref{sec:massless-resum}. 
The Lund plane representation,  depicted in Figs.~\ref{fig:LundPlane-with-mass-1},\ref{fig:LundPlane-with-mass-2},\ref{fig:LundPlane-with-mass-3}, is useful to identify the different integration regions. 

The relevant integration region for the measured jet function is the area shaded in red in Figs.~\ref{fig:LundPlane-with-mass-1},\ref{fig:LundPlane-with-mass-2},\ref{fig:LundPlane-with-mass-3}, which refer to the cases $1-x>\sqrt{\xi},\sqrt{\xi}<1-x<\xi, 1-x>\xi$ respectively. We see that
the jet function receives contributions from both the regions below ($n_f=4$) and above ($n_f=5$) the red dashed lines which marks the heavy quark threshold. For values of $1-x$ smaller than $\sqrt{\xi}$ the $n_f=5$ region freezes at $1-x=\sqrt{\xi}$ and the entire $x$ dependence is given by the $4$-flavour contribution.

We  decompose $j$ as
\begin{align} \label{eq:jet-function-thetas}
j(1-x,\xi)&=j^{(1)}(1-x,\xi)\Theta(1-x -\sqrt{\xi})+j^{(2)}(1-x,\xi)\Theta(\sqrt{\xi}-(1-x)),
\end{align}
and we consider first the case $1-x>\sqrt{\xi}$. 
\begin{figure}[htb]
	\centering
\includegraphics[width=0.8\textwidth,page=3]{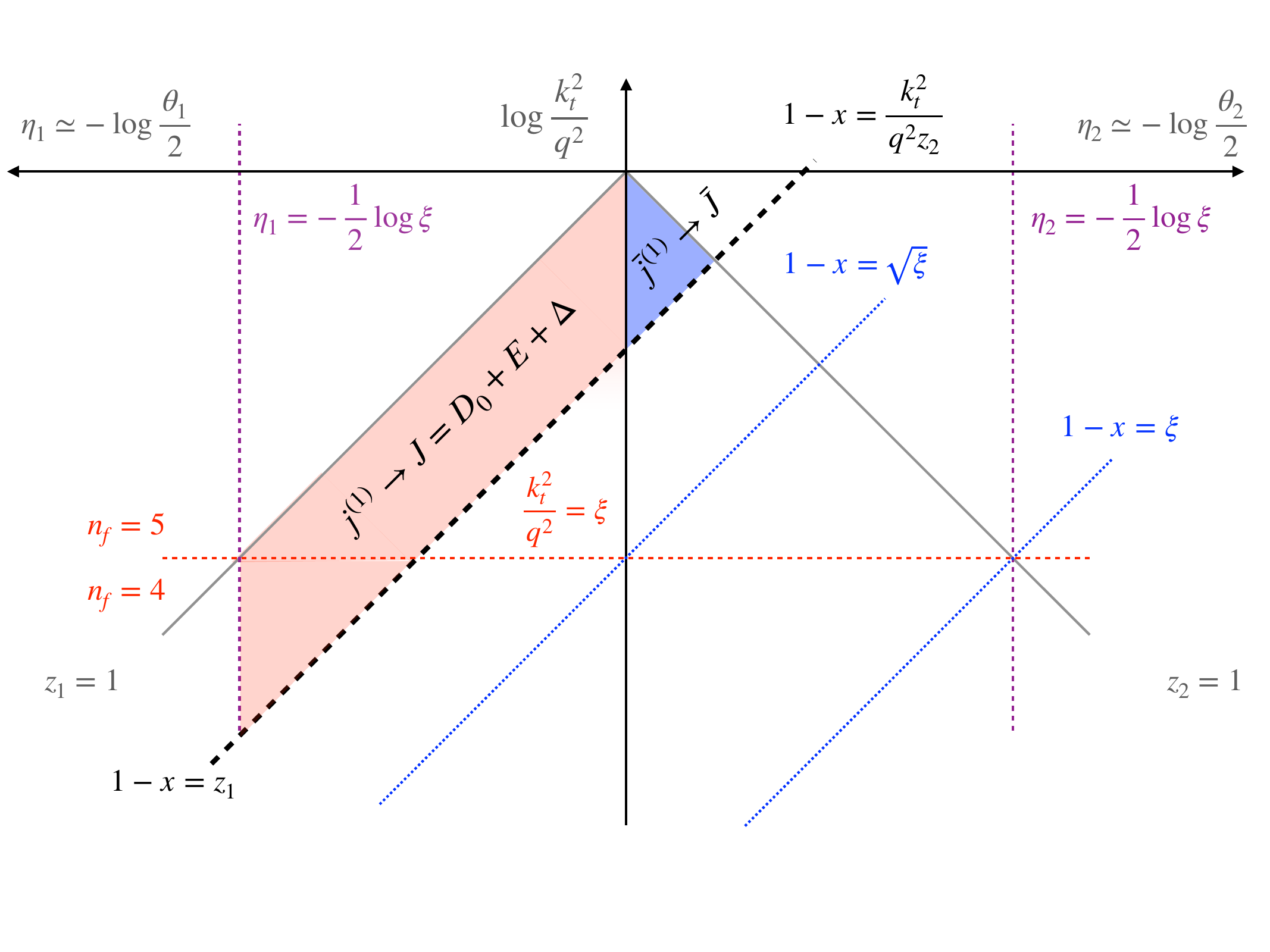}
\vspace{-0.5cm}
	\caption{Lund plane representation of the jet functions.
	The diagonal dashed lines in black represent the value of the observable $1-x$, which is taken to be larger than $\sqrt{\xi}$. The shaded areas indicate the regions of phase space where emissions are vetoed, giving rise to the Sudakov form factor (in red the measured jet function $j$, in blue the recoil jet function $\bar j$.) The corresponding Mellin-space jet functions, $J$ and $\bar{J}$ are also indicated. 
	}
	\label{fig:LundPlane-with-mass-1}
\end{figure}
The $k_t$ integration range splits into two regions, separated by the red dashed line in Fig.~\ref{fig:LundPlane-with-mass-1}, corresponding to the different numbers of active flavours. 
After translating the integration domain from the variables of the Lund representation to the pair $z_1,k_t^2$, we obtain
\begin{equation}\label{eq:J-region1}
	\begin{split}
	j^{(1)}(1-x,\xi)&= \left(-\int^1_{1-x} \de z_1\int^{q^2}_{m^2} \frac{\de k_t^2}{k_t^2}+\int^1_{1-x} \de z_1 \int^{q^2}_{q^2 z_1^2} \frac{\de k_t^2}{k_t^2}\right) \frac{\as^{\text{CMW}}(k_t^2)}{2\pi}P_{gq}(z_1)\\
  &-\int^1_{1-x} \de z_1 \int^{m^2}_{z_1^2 m^2} \frac{\de k_t^2}{k_t^2} \frac{\as^{\text{CMW}}(k_t^2)}{2\pi}P_{gb}(z_1,k_t^2-z_1^2 m^2).
	\end{split}
\end{equation}
Note that in the first line of Eq.~(\ref{eq:J-region1}), which refers to the $n_f=5$ region, we have neglected mass corrections.

Using the definition of the CMW strong coupling, Eq.~(\ref{eq:as_CMW}}), and of the relevant splitting functions, Eqs.~(\ref{eq:masslessAP},\ref{eq:Massive AP}),
it can be checked that this result coincides with $J(N,\xi)$ given in Eq.~(\ref{eq:j-def}), provided that we identify $1/\bar{N}=1-x$, $\mu_0^2=m^2$, and $\mu^2=q^2$. Thus, to NLL we have:
\begin{equation}\label{eq:J1-N}
\tilde j^{(1)}(N,\xi)= j^{(1)}\left(\frac{1}{\bar N},\xi\right)=J(N,\xi).
\end{equation}
In particular, the contribution from the $n_f=4$ region, i.e.\ the second line of Eq.~(\ref{eq:J-region1}), corresponds to the initial condition $D_0$ for the fragmentation function, while the $n_f=5$ contribution, i.e.\ the first line of Eq.~(\ref{eq:J-region1}), corresponds to the NLL contributions to $E+\Delta$.

As a side comment, it is interesting to note that the fragmentation function evolved up to the hard scale, i.e.\ the factor $\exp(E+D_0)$, can be obtained from Eq.~(\ref{eq:jet-function-J}),  provided the integration is extended to the region bounded by the lines $z_1=1-x$, $z_1=1$, $\eta_1=-\frac{1}{2}\log\xi$, $\log\frac{k_t^2}{q^2}=0$, thereby including not only the region where the emission is collinear to the antiquark momentum (shaded in blue in Fig.~\ref{fig:LundPlane-with-mass-1}),  but also part of the region outside the kinematic limit $z_2=1$. The role of the factor  $\exp\Delta$ is therefore to subtract off this unphysical behaviour of the soft part of the fragmentation function.~\footnote{We thank Gavin Salam for discussions on this topic.}

This calculation confirms the result of Eq.~(\ref{eq:Jnspace_region1}), namely the fact that the leading logarithms originating from the second and the third terms in Eq.~(\ref{eq:J-region1}) do cancel at first order but fail to do so at higher orders because the running coupling receives contributions from different flavour regions, i.e.\ from above and below the $\frac{k_t^2}{q^2} = \xi$ line on the Lund plane. Furthermore, the analysis in momentum space allows us to identify the region in which this happens in terms of physical quantities, $1-x >\sqrt{\xi}$, as opposed to $\xi \bar N^2<1$ found in Eq.~(\ref{eq:Jnspace_region1}).

We now turn to the case $1-x<\sqrt{\xi}$, illustrated in Fig.~\ref{fig:LundPlane-with-mass-2}.
We find
\begin{align} \label{eq:J-region2}
		j^{(2)}(1-x,\xi)=&\left(-\int^1_{\sqrt{\xi}} \de z_1\int^{q^2}_{m^2} \frac{\de k_t^2}{k_t^2}+\int^1_{\sqrt{\xi}} \de z_1 \int^{q^2}_{q^2 z_1^2} \frac{\de k_t^2}{k_t^2}\right) \frac{\as^{\text{CMW}}(k_t^2)}{2\pi}P_{gq}(z_1)\\ \notag
		+&\left(-\int^1_{1-x} \de z_1 \int^{m^2}_{z_1^2 m^2} \frac{\de k_t^2}{k_t^2}+\int^{\sqrt{\xi}}_{1-x} \de z_1 \int^{m^2}_{q^2 z_1^2} \frac{\de k_t^2}{k_t^2}\right) \frac{\as^{\text{CMW}}(k_t^2)}{2\pi}P_{gb}(z_1,k_t^2-z_1^2m^2).
\end{align}
As anticipated, the $n_f=5$ contribution (first line) is now evaluated at the transition $1-x=\sqrt{\xi} $ and bears no $x$ dependence. Furthermore, by inspecting the second line, we confirm that leading (double) logarithms in $1-x$ cancel between the two contributions to all orders, as already found in the Mellin space computation, Eq.~(\ref{eq: J without LL}).
\begin{figure}[htb]
	\centering
\includegraphics[width=0.8\textwidth,page=4]{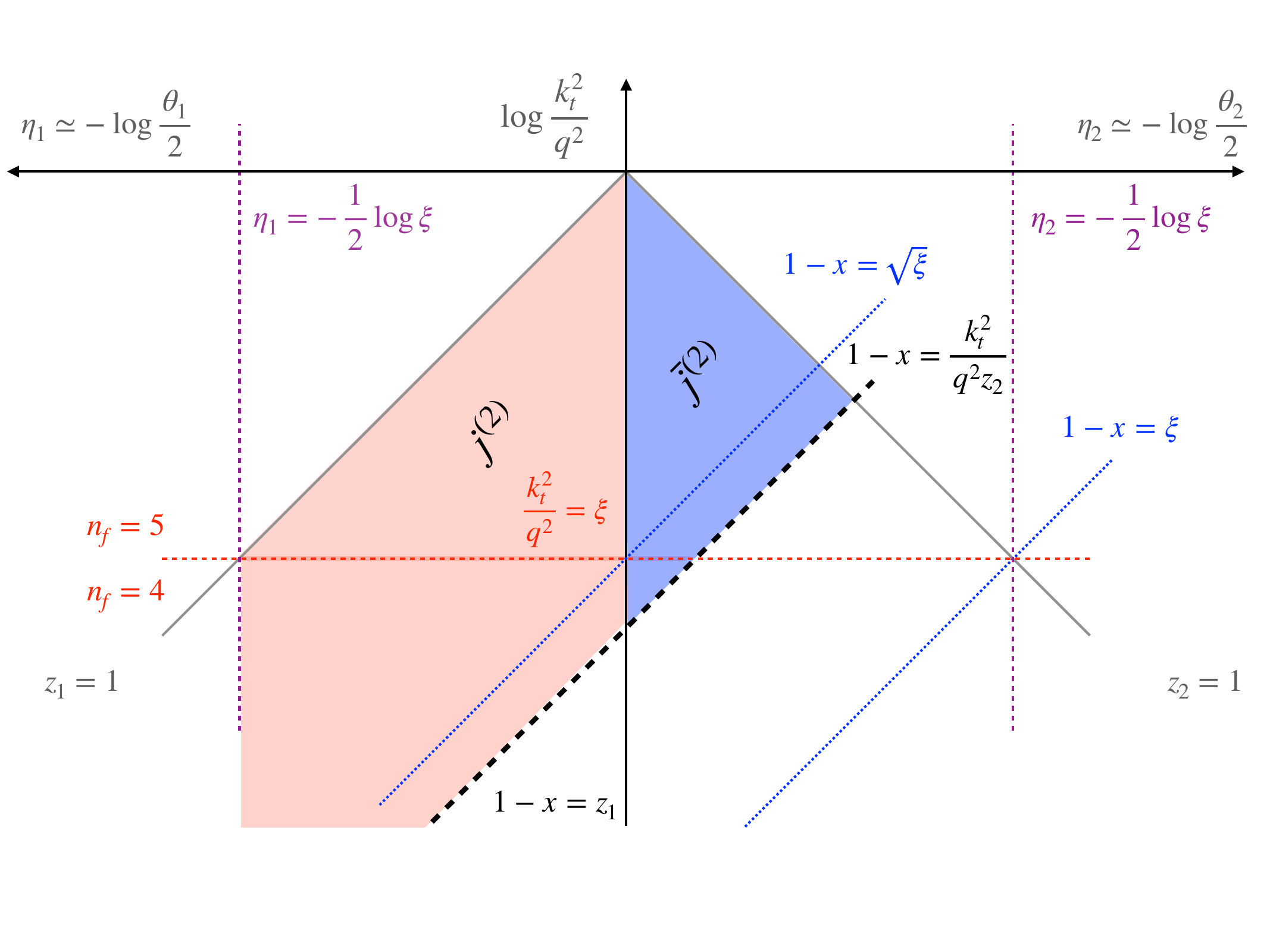}
\vspace{-0.5cm}
	\caption{Same as Fig.~\ref{fig:LundPlane-with-mass-1}, for $\xi < 1-x <\sqrt{\xi}$.
	}
	\label{fig:LundPlane-with-mass-2}
\end{figure}

We also note that, in this case, double logarithms of the mass appear. Similarly to what happens in the region $1-x>\sqrt{\xi}$ for the double logarithms of $1-x$ (or $N$), this contribution cancels at $\order{\as}$ but not at higher orders, because of the different evolution of the running coupling in different flavour regions. 
Finally, we note that there is no transition for $j(1-x,\xi)$ associated to $1-x=\xi$.
As in the previous case,  the result in Mellin space  in the region $1-x<\sqrt{\xi}$ to NLL accuracy is given by
\begin{equation}\label{eq:J1-N}
\tilde j^{(2)}(N,\xi)= j^{(2)}\left(\frac{1}{\bar N},\xi \right).
\end{equation}
These results are also in agreement with the fragmentation function analysis of Section~\ref{sec:massless-resum}. 
\begin{figure}[htb]
	\centering
\includegraphics[width=0.8\textwidth,page=5]{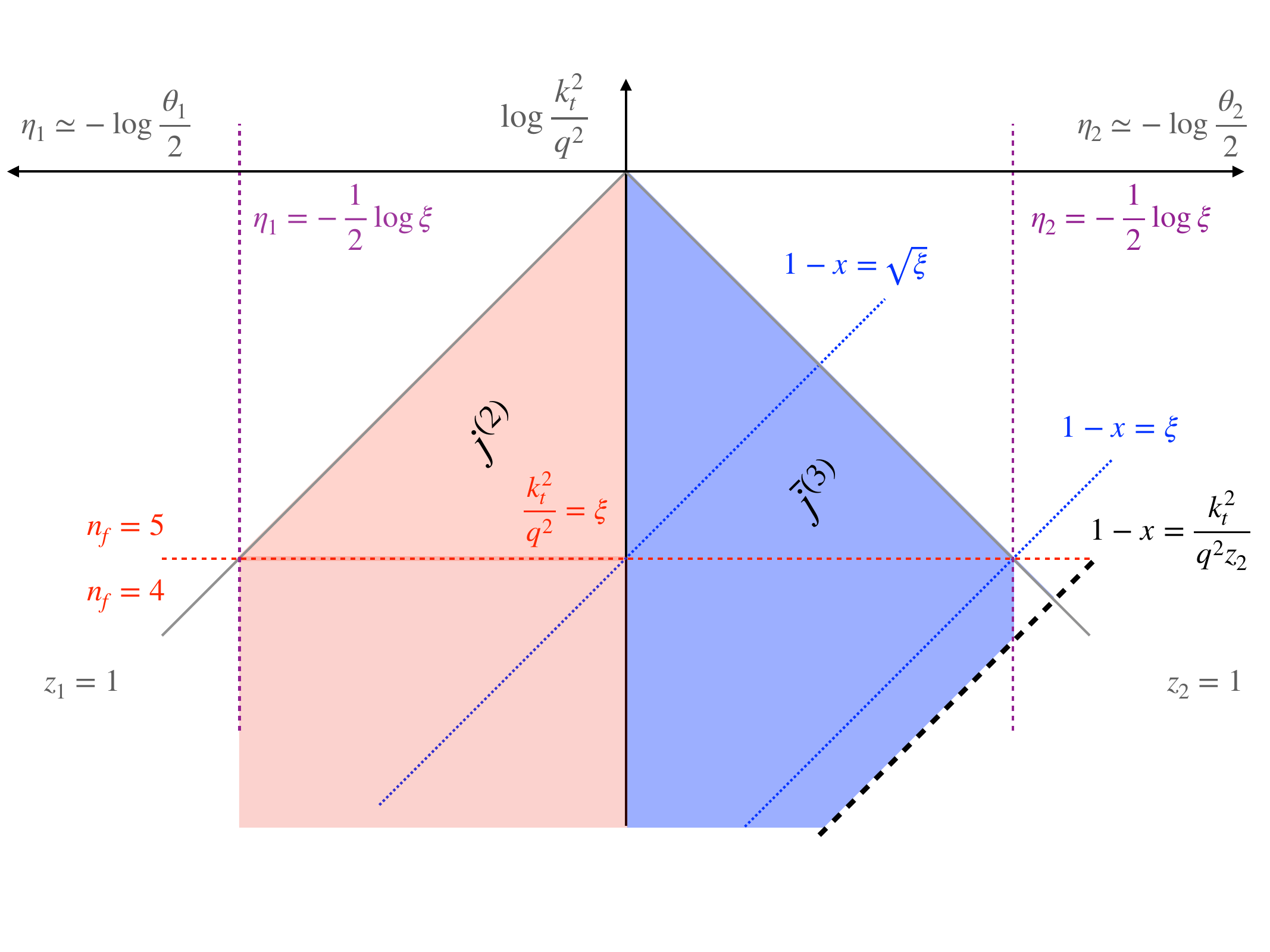}
\vspace{-0.5cm}
	\caption{Same as Fig.~\ref{fig:LundPlane-with-mass-1}, for $1-x <\xi$.
	}
	\label{fig:LundPlane-with-mass-3}
\end{figure}

\subsubsection{Calculation of the unmeasured jet function}\label{sec:jbar-lund}
In this section we focus on the computation of $\bar{j}(1-x,\xi)$ given in Eq~(\ref{eq:jet-function-Jbar}).  In this case there is a second transition at $1-x=\xi$, which was not present in the case of the measured jet function calculation. This transition is of kinematic origin and can be understood already at fixed coupling. With this assumption, the integrals in Eq.~(\ref{eq:jet-function-Jbar}) are straightforward and the result reads
\begin{align} \label{eq:Jbar fix coupling}
		\text{if}\quad  1-x> \xi, \quad \bar{j}(1-x,\xi)&=\frac{\as\cf}{\pi}\left(-\frac{1}{2}\log^2{(1-x)}-\frac{3}{4}\log{(1-x)}\right),\\
		\text{if}\quad 1-x<\xi, \quad \bar{j}(1-x,\xi)&=\frac{\as\cf}{\pi}\left(\frac{1}{2}\log^2{\xi}-\log{(1-x)}\log{\xi}-\log{(1-x)}+\frac{1}{4}\log{\xi}\right). \nonumber
\end{align}	
Remarkably, after identifying $1-x=\frac{1}{\bar{N}}$, the two expressions above reproduce the logarithmic terms present in the expansion at $\order{\as}$ of the resummed formula in massless and massive schemes,  Eqs.~(\ref{fragm-master}) and~(\ref{eq:massive resummation-bis}), respectively. Specifically, when $1-x>\xi$, $\bar{j}$ reproduces the double logarithmic behaviour of Eq.~(\ref{fragm-master}) at $\order{\as}$. On the other hand,  $\bar j$ captures both the single logarithms of Eq.~(\ref{eq:massive resummation-bis}) and the mass double logarithms of  Eq.~(\ref{eq:Cone}) when $1-x<\xi$. We also notice that at the transition point $1-x=\xi$, $\bar{j}$ is continuous.
Therefore, the momentum space calculation of $\bar j$ allows us to better understand and resolve the issue of the non-commutativity of the $x\to 1$ and $\xi \to 0$ limits. Indeed, by computing the recoiling jet function in the quasi-collinear limit, rather than in the massless collinear limit (as done, for instance, in~\cite{Cacciari:2001cw}), we are able to obtain an expression that interpolates between Eqs.~(\ref{fragm-master}) and~(\ref{eq:massive resummation-bis}). 
Furthermore, we are also able to confirm that the double logarithmic terms in Eq.~(\ref{eq:Cone}) are entirely generated by $\bar{j}$ and the mass that appears is the one of the undetected antiquark $\bar{b}$. 
Note that this effect is entirely driven by the behaviour of the observable in the hemisphere collinear to the antiquark, and it is absent on the other side of the Lund plane.

The above considerations can be generalised to NLL by including running coupling corrections (see also~\cite{Aglietti:2006wh,Hoang:2018zrp}). We have
\begin{align} \label{eq:recoil-jet-function-thetas}
\bar j(1-x,\xi)&=\bar j^{(1)}(1-x,\xi)\Theta(1-x -\sqrt{\xi})+\bar j^{(2)}(1-x,\xi)\Theta(\sqrt{\xi}-(1-x))\Theta(1-x-\xi)
\nonumber \\ &+\bar j^{(3)}(1-x,\xi)\Theta(\xi-(1-x)).
\end{align}
If $1-x>\sqrt{\xi}$ (see Fig.~\ref{fig:LundPlane-with-mass-1})  the recoil jet function is computed with $n_f=5$ active flavours, and the quark mass neglected.
Hence
	  \begin{equation}\label{eq:Jbar_FF}
	  	\bar{j}^{(1)}(1-x,\xi)=\bar{j}^{(1)}(1-x)= -\int^1_{1-x}\de z_2 \int^{z_2q^2}_{z_2^2 q^2} \frac{\de k_t^2}{k_t^2} \frac{\as^{\text{CMW}}(k_t^2)}{2\pi} P_{gq}(z_2).
\end{equation}
Thus, the calculation of $\bar j$ coincides with the one performed in Mellin space in the massless case, Eq.~(\ref{eq: barJ}), after the replacement  $1-x=\frac{1}{\bar{N}}$:
\begin{equation}\label{eq:J1bar-N}
\tilde{\bar j}^{(1)}(N)= j^{(1)}\left(\frac{1}{\bar N}\right)=\bar J(N).
\end{equation}
 
Lowering the value of $1-x$, we enter the intermediate region $\xi<1-x<\sqrt{\xi}$,  where the $x$ dependence arises both from the $n_f=5$ and from the $n_f=4$ contributions, as seen by inspection of the Lund diagram in Fig~\ref{fig:LundPlane-with-mass-2}. 
Therefore, in the intermediate region we find 
\begin{align} \label{eq:Jbar_intermediate_region}
&  		\bar{j}^{(2)}(1-x,\xi)=
	\bar{j}^{(1)}(\sqrt{\xi})
		-\int^{m^2}_{q^2(1-x)^2}\frac{\de k_t^2}{k_t^2}\frac{\as^{\text{CMW}}(k_t^2)}{2\pi} \int^{\frac{k_t^2}{q^2(1-x)}}_{\frac{k_t}{q}}\de z_2\, P_{gq}(z_2) \nonumber \\ &-\int^{q^2(1-x)}_{m^2}\frac{\de k_t^2}{k_t^2}\frac{\as^{\text{CMW}}(k_t^2)}{2\pi} \int^{\frac{k_t^2}{q^2(1-x)}}_{\frac{k_t^2}{q^2\sqrt{\xi}}}\de z_2\, P_{gq}(z_2)
  		-\int^{q^2\sqrt{\xi}}_{q^2(1-x)} \frac{\de k_t^2}{k_t^2} \frac{\as^{\text{CMW}}(k_t^2)}{2\pi}\int^{1}_{\frac{k_t^2}{q^2 \sqrt{\xi}}} \de z_2\, P_{gq}(z_2), 
  	\end{align}
where  the first term is Eq.~(\ref{eq:Jbar_FF}), evaluated at the transition $1-x=\sqrt{\xi}$.	In this case we found it convenient to swap the order of the integrations in $z_2$ and $k_t^2$. In this way, the $z_2$ integrals are all of the same form; the result, to NLL accuracy, has the form $c_0 + c_1 \log{\frac{k_t^2}{q^2}}$ with $c_0$ and $c_1$
    independent of $k_t^2$, but dependent on $x,q^2,\xi$ through the integration bounds.
 We note that the first and second integrals receive logarithmic contributions only by the soft-collinear part of the splitting function, i.e.\ we can take in those integrals $P_{gq}=\frac{2}{z_2}$, while in the third one, the upper limit of the $z_2$ integration is 1 and therefore the complete massless splitting function is needed. 
Finally, we must consider the region $1-x<\xi$. We find
\begin{align}\label{eq:Jbar-final-region}
		\bar{j}^{(3)}(1-x,\xi)&= \bar{j}^{(2)}(\xi,\xi)
		-\int^{m^2\xi}_{q^2(1-x)^2} \frac{\de k_t^2}{k_t^2} \frac{\as^{\text{CMW}}(k_t^2)}{2\pi}\int^{\frac{k_t^2}{q^2 (1-x)}}_{\frac{k_t}{q}} \de z_2 \, P_{gq}(z_2)\\
		&-\int^{\frac{q^2(1-x)^2}{\xi}}_{m^2\xi}\frac{\de k_t^2}{k_t^2}\frac{\as^{\text{CMW}}(k_t^2)}{2\pi}\int^{\frac{k_t^2}{m^2}}_{\frac{k_t^2}{q^2(1-x)}} \de z_2 \, P_{gq}(z_2)  \nonumber \\ &
		-\int^{m^2}_{\frac{q^2(1-x)^2}{\xi}}\frac{\de k_t^2}{k_t^2}\frac{\as^{\text{CMW}}(k_t^2)}{2\pi}\int^{\frac{k_t}{m}}_{\frac{k_t^2}{m^2}}\de z_2 \, P_{gb}(z_2,k_t^2-z^2m^2). \nonumber
\end{align}
The first contribution does not depend on $x$ and it is given by the result of the previous region, Eq.~(\ref{eq:Jbar_intermediate_region}), evaluated at $1-x=\xi$. The $x$-dependent contributions arise from the $n_f=4$ region, as shown in Fig.~\ref{fig:LundPlane-with-mass-3}. This term is sensitive to finite mass corrections, as discussed at the beginning of this section. In order to perform the integrals we adopted the same strategy explained above.  

Mellin space results in each region are obtained by the procedure adopted in the previous cases:
\begin{equation}\label{eq:Jbar-N}
\tilde{\bar j}^{(i)}(N,\xi)= j^{(i)}\left(\frac{1}{\bar N},\xi \right), \quad i=1,2,3.
\end{equation}
Explicit results are given in App.~\ref{app:calculations}.

\section{All-order matching and numerical results}\label{sec:results_decay}

We can now exploit the results of the previous section to combine, in a consistent way, 
the NLL resummed result differential rate for the process in Eq.~(\ref{eq:Hdecay}) in the fragmentation-function approach, $\widetilde{\Gamma}_{\ell=1, \ell_2=1}^{(5,\text{res})}$,  with the resummed calculation $ \widetilde{\Gamma}^{(4,\text{res})}_{\ell_1=0}$ in the massive scheme. To this purpose, we must take into account that constant (i.e.\ $N$ independent) terms are not included in the momentum space results; they must therefore be recovered by matching with the calculations presented in Section~\ref{sec:decay}.
We have
\begin{equation}\label{eq:final-result}
\frac{1}{\Gamma_0}\frac{\de \Gamma}{\de x}= \int_{c-i \infty}^{c+ i \infty}\frac{\de N}{2 \pi i}\, x^{-N}
\begin{cases} 	
\widetilde{\Gamma}^{(1)}(N,\xi), & \text{if}\; 1-x >\sqrt{\xi},\\
\widetilde{\Gamma}^{(2)}(N,\xi), & \text{if}\; \xi<1-x <\sqrt{\xi}, \\
\widetilde{\Gamma}^{(3)}(N,\xi), & \text{if} \; 1-x <\xi,\\
\end{cases}	
\end{equation}
with
\begin{align}\label{eq:NLL2}
\widetilde{\Gamma}^{(1)}(N,\xi)&=\widetilde{\Gamma}_{\ell=1, \ell_2=1}^{(5,\text{res,sub})} \exp  \left[\tilde{j}^{(1)}(N,\xi)+\tilde{\bar{j}}^{(1)}(N)\right], \nonumber \\
\widetilde{\Gamma}^{(2)}(N,\xi)&= \widetilde{\Gamma}^{(\text{match})} \exp \left[\tilde{j}^{(2)}(N,\xi)+\tilde{\bar{j}}^{(2)}(N,\xi)\right], \nonumber \\
\widetilde{\Gamma}^{(3)}(N,\xi)&= \widetilde{\Gamma}^{(4,\text{res,sub})}_{\ell_1=0}  \exp \left[\tilde{j}^{(2)}(N,\xi)+\tilde{\bar{j}}^{(3)}(N,\xi)\right].
\end{align}
In the three expressions above,  the suffix sub denotes the subtracted version of the 5- and 4-flavour calculations, that is, the factors that do not contain logarithms of $N$.
In particular, 
\begin{align}
	\widetilde{\Gamma}_{\ell=1, \ell_2=1}^{(5,\text{res-sub})}(N,\xi)&=
				 \left(1+ \frac{\as(\mu^2) \cf}{\pi}\mathcal{C}_0^{(1)} \right) \left(1+ \frac{\as(\mu_0^2) \cf}{\pi}\mathcal{D}_0^{(1)} \right) \mathcal{E}^{(\text{sub})}(N,\mu_0^2,\mu^2),
\end{align}
where $\mathcal{E}^{(\text{sub})}(N,\mu_0^2,\mu^2)$ is defined in Eq.~(\ref{eq:DGLAPsubtracted}). 
Similarly, 
\begin{align}
		\widetilde{\Gamma}^{(4,\text{res-sub})}_{\ell_1=0}(N,\xi)&= \left(1+ \frac{\as(\mu_0^2)}{\pi}\mathcal{K}^\text{sub}_1(\xi)\right) \exp \left[-2\, \gszerotilde(\beta) \int^1_{1/\bar N} \frac{\de z}{z}   \frac{\as\left(z^2 \mu_0^2\right)}{\pi}\right],
\end{align}
where $\mathcal{K}_1^\text{sub}=\mathcal{K}_1-\frac{\cf}{2}\left(\log^2\xi-\log\xi\right)$. 

By construction, $\widetilde{\Gamma}^{(1)}(N,\xi)$ coincides with the 5-flavour result of~\cite{Cacciari:2001cw}.
We note that $\widetilde{\Gamma}^{(3)}(N,\xi)$ is fully consistent, to NLL accuracy, with $\widetilde{\Gamma}^{(4,\text{res})}_{\ell_1=0}$ given in 
Eq.~(\ref{eq:massive resummation-bis}), but differs from it in three respects.
First, the argument of the running coupling in $\widetilde{\Gamma}^{(3)}(N,\xi)$ was chosen to be $\mu_0^2$, which is of the order of the heavy quark mass, while it is chosen to be $\mu^2 \simeq q^2$ in Eq.~(\ref{eq:massive resummation-bis}). The difference is subleading; the choice $\mu_0^2$ appears to be more natural for a 4-flavour calculation.
Second, the momentum-space calculation of the jet functions is performed with the strong coupling at two loops, in the CMW scheme. 
Third, all logarithms of $\xi$ are now exponentiated in the jet functions.

The contribution  $\widetilde{\Gamma}^{(2)}(N,\xi)$ performs the all-order matching between the two results. Note that, while our calculation determines all NLL contributions (both in $\xi$ and $\bar N$), it does not determine the matching constant $\widetilde{\Gamma}^{(\text{match})}$. 
Furthermore, because in this intermediate region $1-x<\sqrt{\xi}$, we cannot simply match to the massless calculation, as done in the first region.
However, we have already noted (see Eq.~(\ref{eq:J-region2}) and below) that in this region, the 5-flavour contribution to the jet function is frozen and evaluated at the transition $1-x=\sqrt{\xi}$, which in Mellin space corresponds to $\bar N^{-1}=\sqrt{\xi}$. Therefore, we find it natural to extend this to the full DGLAP contribution. This can be achieved by choosing:
\begin{equation} \label{eq:matching-choice}
\widetilde{\Gamma}^{(\text{match})}={\widetilde{\Gamma}_{\ell=1, \ell_2=1}^{(5,\text{res,sub})}}|_{\bar N^{-1}=\sqrt{\xi}}.
\end{equation}
Note that with this choice, the result in the second region is consistent with the calculation of the fragmentation function with flavour threshold described in Sect.~\ref{sec:massless-resum}.
More refined matching options are possible. For instance, we could require continuity at both $\bar N^{-1} = \sqrt{\xi}$ and $\bar N^{-1} = \xi$ and use an interpolating function (see e.g.~\cite{Aglietti:2022rcm}). We leave detailed studies of the numerical impact of different matching choices to future work.

\begin{figure}
	\centering
\includegraphics[width=0.48\textwidth]{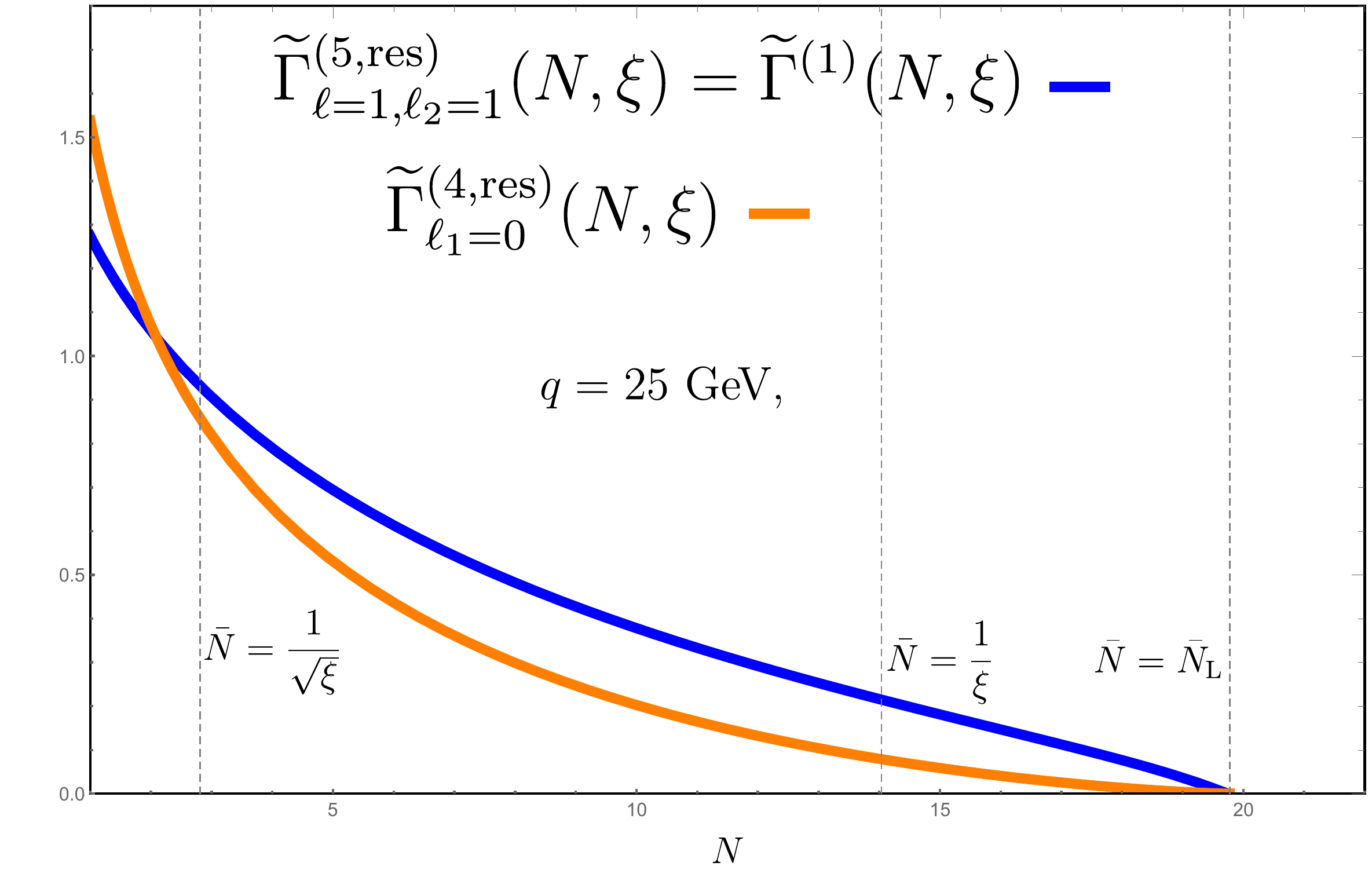}
\includegraphics[width=0.48\textwidth]{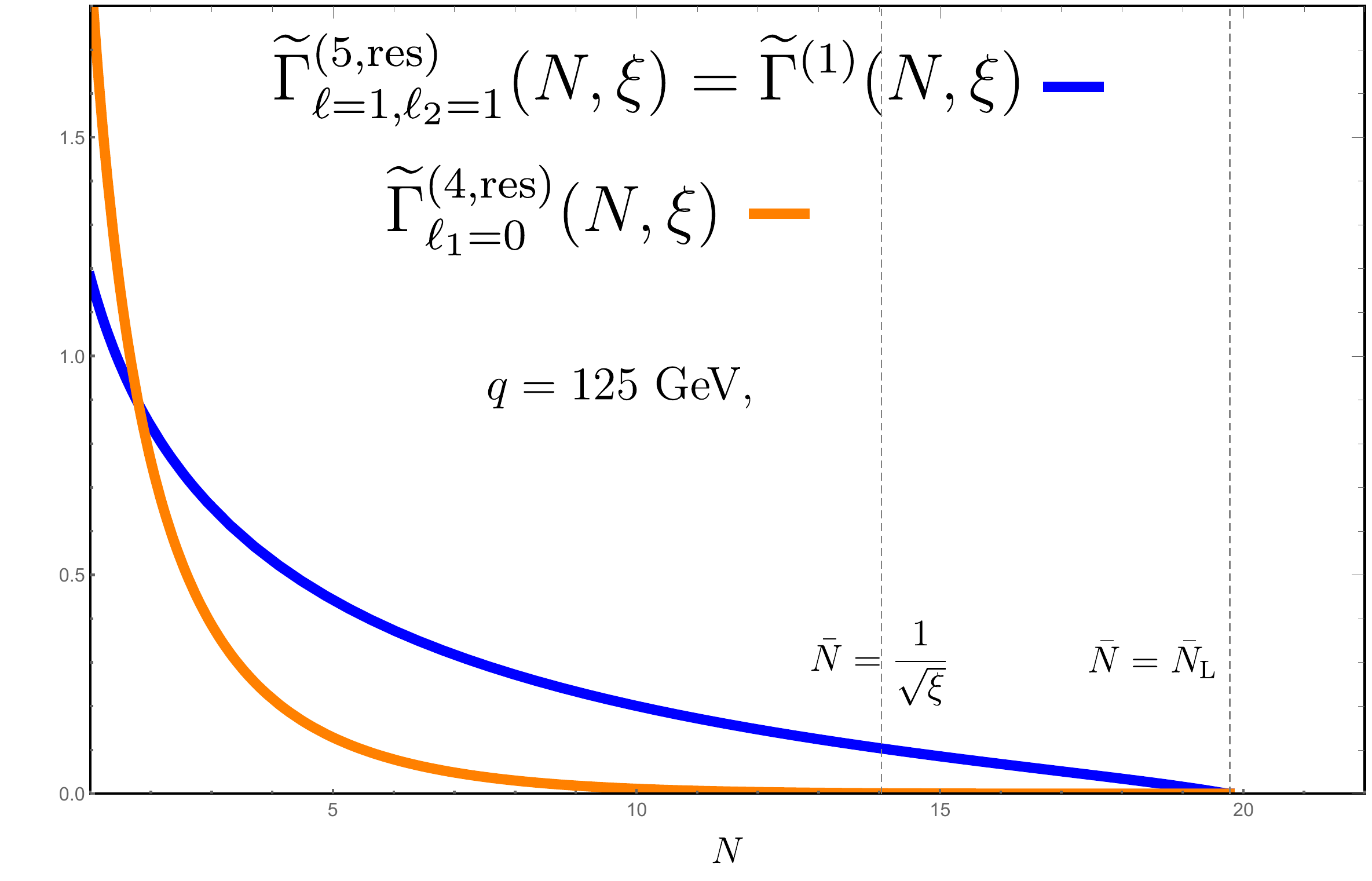}
\includegraphics[width=0.48\textwidth]{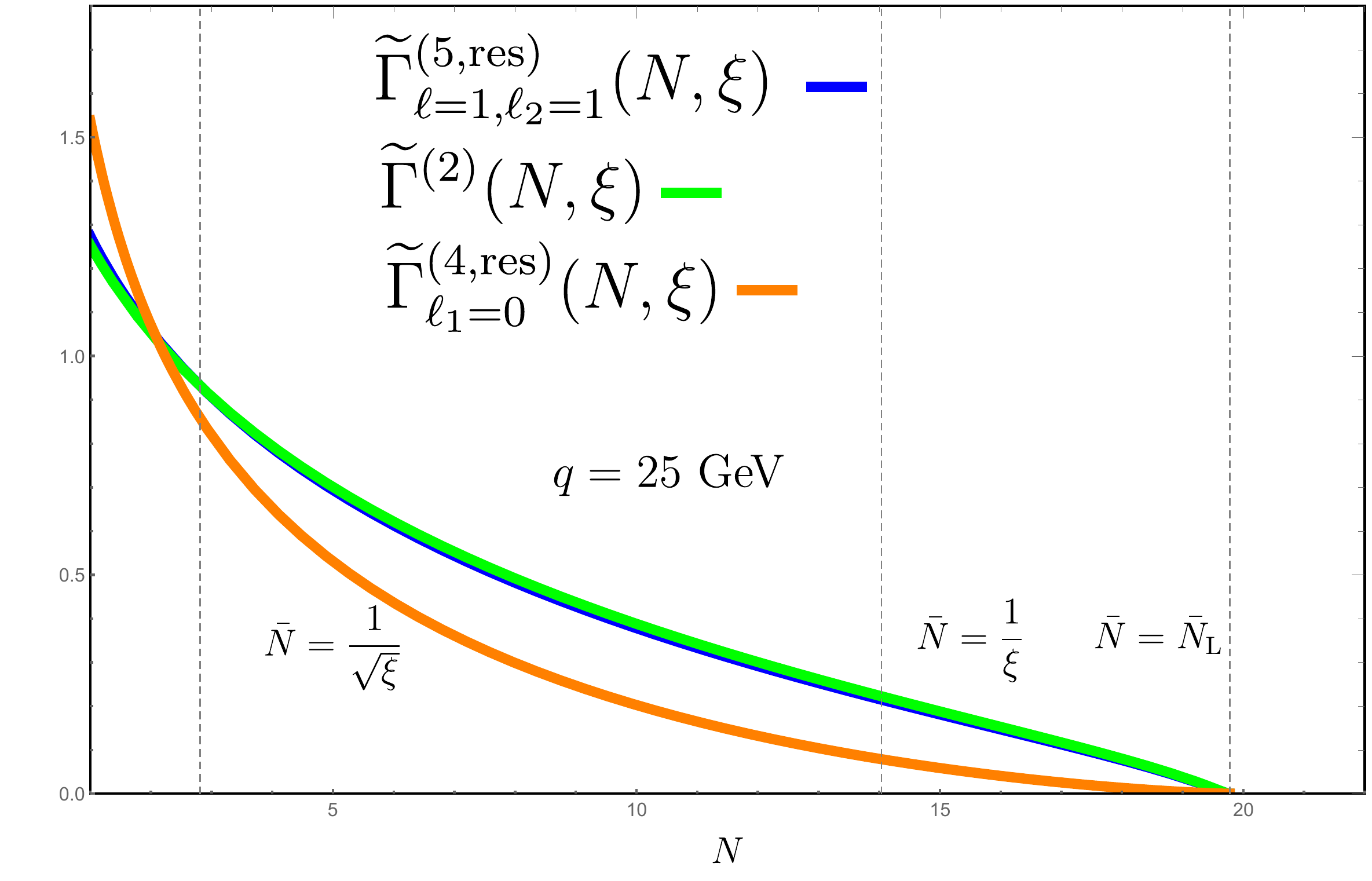}
\includegraphics[width=0.48\textwidth]{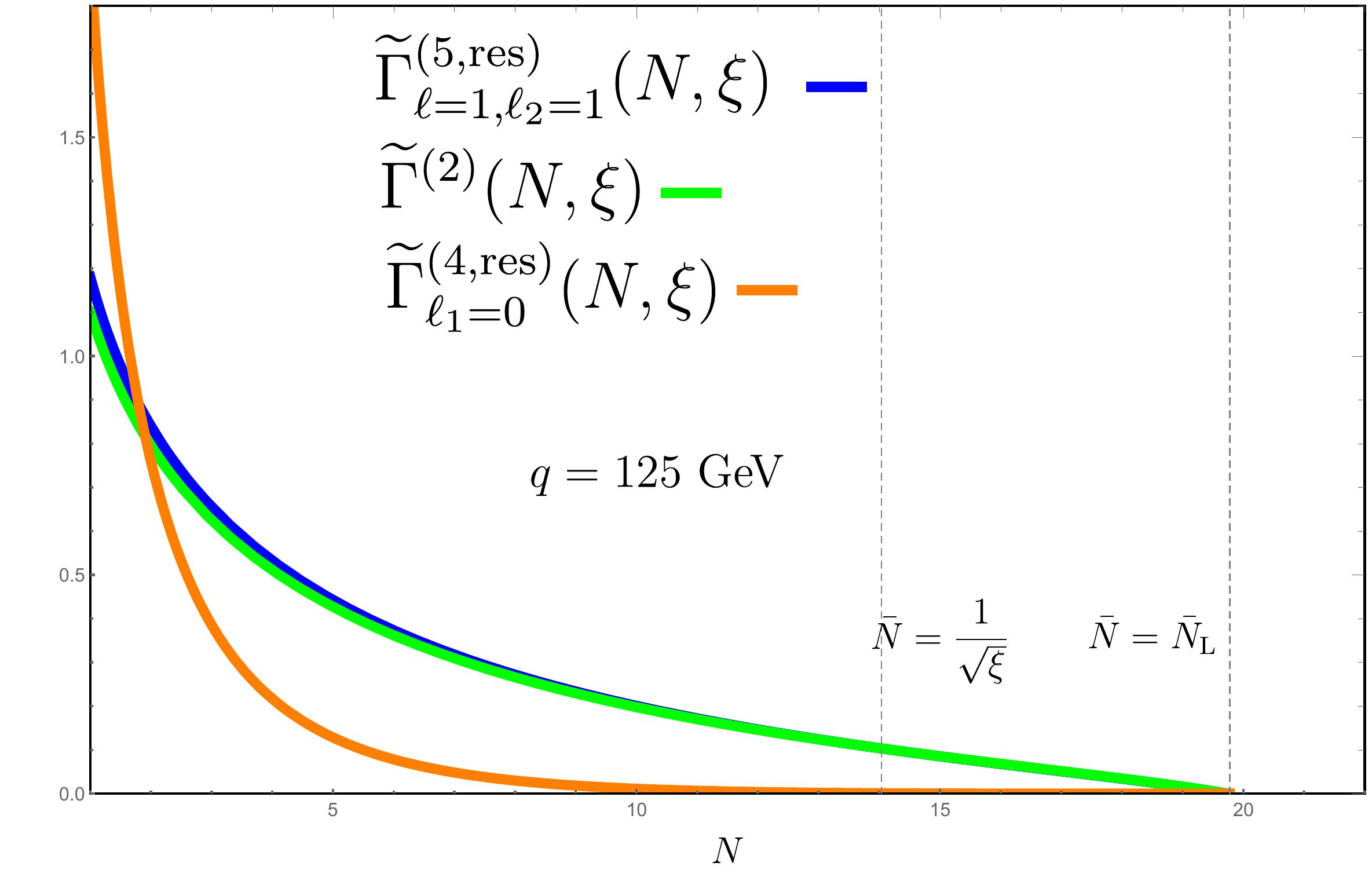}
\includegraphics[width=0.48\textwidth]{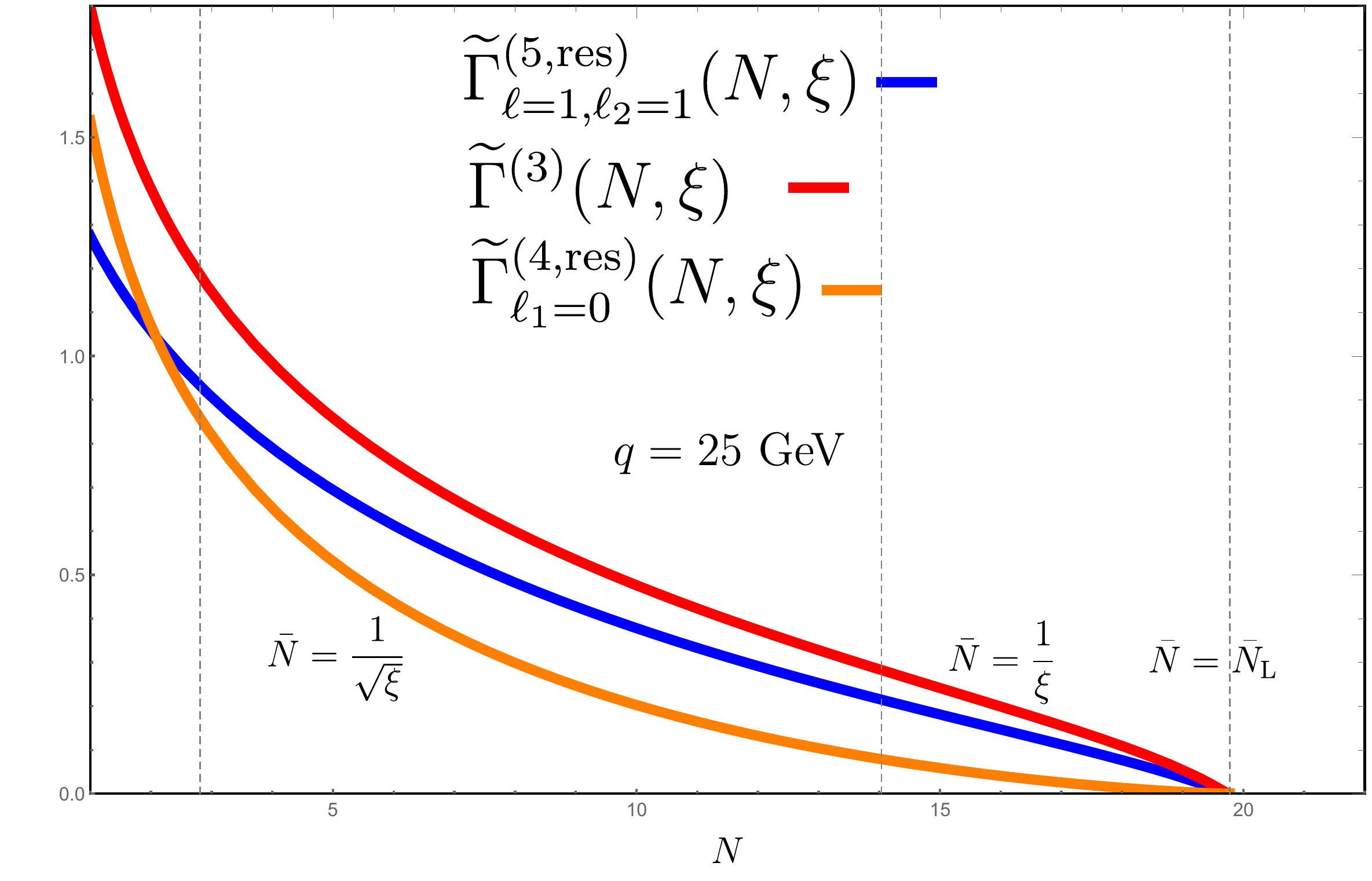}
\includegraphics[width=0.48\textwidth]{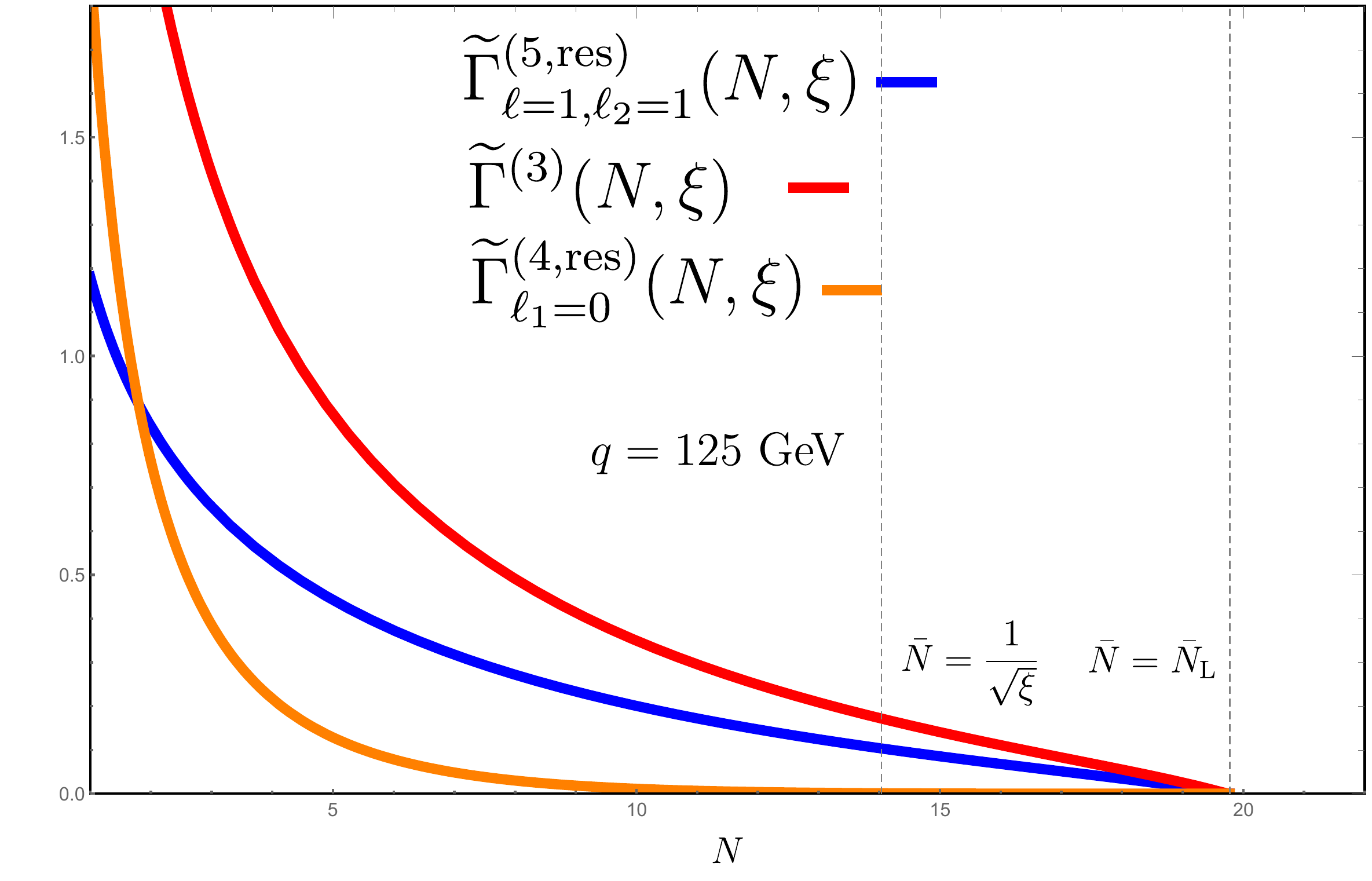}
	\caption{Our results, Eq.~(\ref{eq:NLL2}), plotted as a function of $N$, on the real axis, for two different values of the hard scale. Plots on the left are for $q=25$~GeV, while the ones on the right for $q=125$~GeV. In both cases we set have $\mu_0^2=\muOf^2=m^2$, and $\mu^2=\muf^2=q^2$. The vertical lines mark the values of $N$ that corresponds to the transitions in momentum space, i.e.\ $\bar N^{-1}\sim 1-x=\sqrt{\xi}$, and $\bar N^{-1}\sim 1-x=\xi$, and the position of the Landau pole. Note that for the 125~GeV case, $\xi^{-1}> \bar N_{\text L}$.
	}
	\label{fig:results-different-regions}
\end{figure}

The three functions $\Gamma^{(1)},\Gamma^{(2)},\Gamma^{(3)}$ are shown in Fig.~\ref{fig:results-different-regions} as functions of $N$, taken to be real and positive, 
for $m=5$ GeV
and two different values of $q$, namely $q=25$~GeV, on the left, and $q=125$~GeV, on the right.
On each plot, we also indicate three vertical lines. Two of them mark the values of $N$ that corresponds to the transitions in momentum space, i.e.\ $\bar N^{-1}\sim 1-x=\sqrt{\xi}$
and $\bar N^{-1}\sim 1-x=\xi$. The third line indicates the position of the Landau pole of the strong coupling,
\begin{equation}
\bar N_\text{L} =e^\frac{1}{2 \as(m^2) \beta_0^{(4)}}.
\end{equation}
Resummed formulae are not reliable beyond this value.
Note that the vertical line  corresponding to $\bar N=\frac{1}{\xi}$ does not appear on the plots for $q=125$~GeV, because in this case $\xi^{-1}> \bar N_\text{L}$, 
We also compare to the results obtained in the massless scheme, Eq.~(\ref{eq:FF-sep}), with $\mu^2=\muf^2=q^2$, and $\mu_0^2=\muOf^2=m^2$ (in blue), and in the massive scheme, Eq.~(\ref{eq:massive resummation-bis}), but with $m^2$ as the running coupling scale (in orange). 

Let us comment the results in the different regions. 
We start by the functions to be used in the region $1-x>\sqrt{\xi}$, which are shown in the top plots of Fig.~\ref{fig:results-different-regions}.
Because of the choice of the matching function, Eq.~(\ref{eq:matching-choice}), we have that $\widetilde{\Gamma}^{(1)}=\widetilde{\Gamma}_{\ell=1, \ell_2=1}^{(5,\text{res})}$. 
Thus, in this region our result coincides with the one of Ref.~\cite{Cacciari:2001cw}.
The massive result instead differs substantially and the difference because more pronounced at larger $q^2$ because of the presence of un-resummed logarithms.

The plots in the middle instead show our results for region 2 (in green), compared to the results present in literature. Here, in principle, our result differs from the massless calculation of Ref.~\cite{Cacciari:2001cw} because of the treatment of the heavy-quark threshold in the running coupling. However, this difference turns out to be remarkably small, irrespectively of the value of $\xi$ and $N$, as demonstrated by the fact that the blue and green curves largely overlap.

Finally, in the third region (bottom plots) our result differs from both massless and massive calculations. While the former is not surprising because we have performed the calculation in the quasi-collinear limit, the latter is due to the exponentiation of the double logarithmic mass contributions discussed in Eq.~(\ref{eq:Cone}).

\begin{figure}
	\centering
\includegraphics[width=0.48\textwidth]{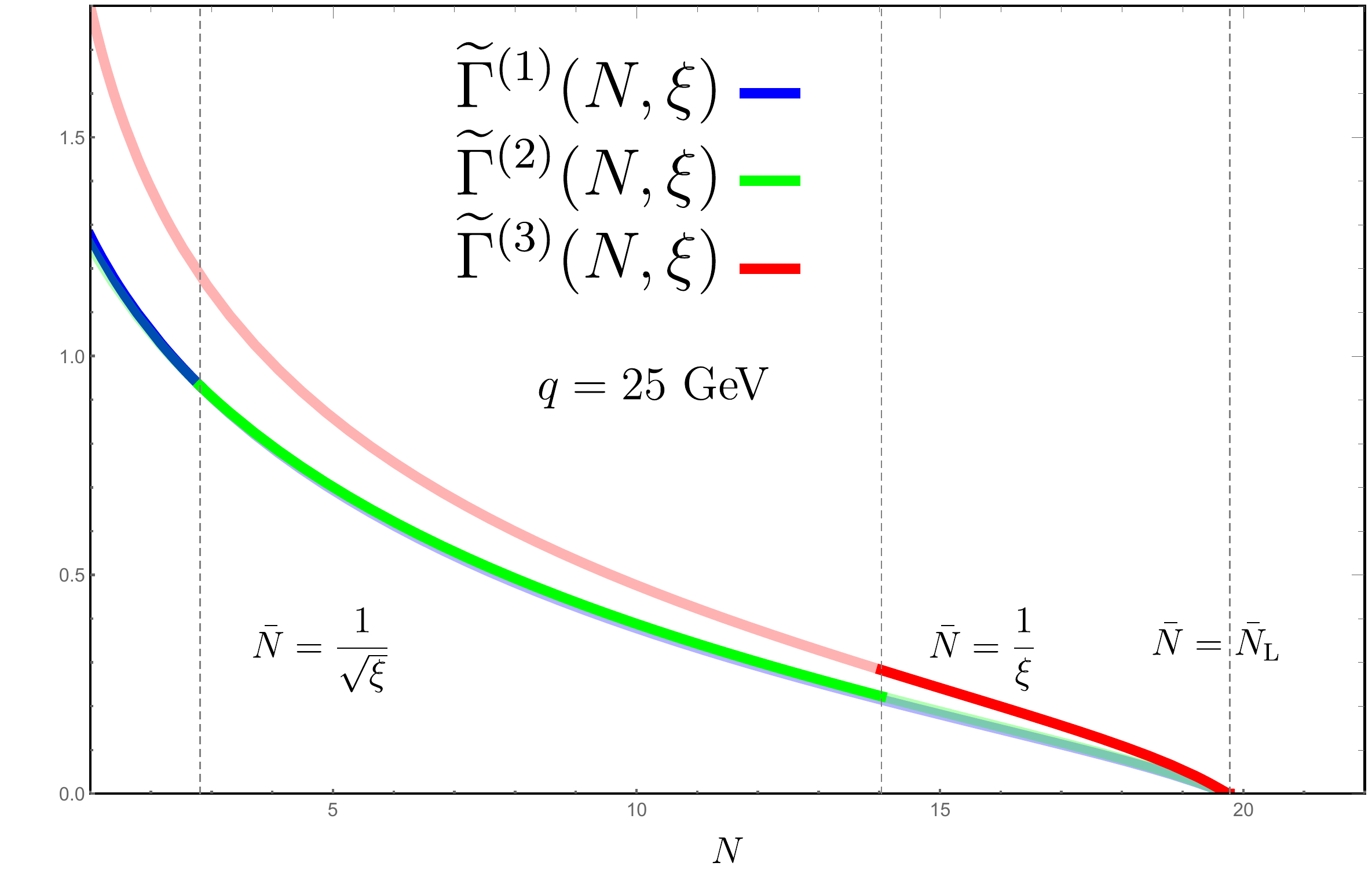}
\includegraphics[width=0.48\textwidth]{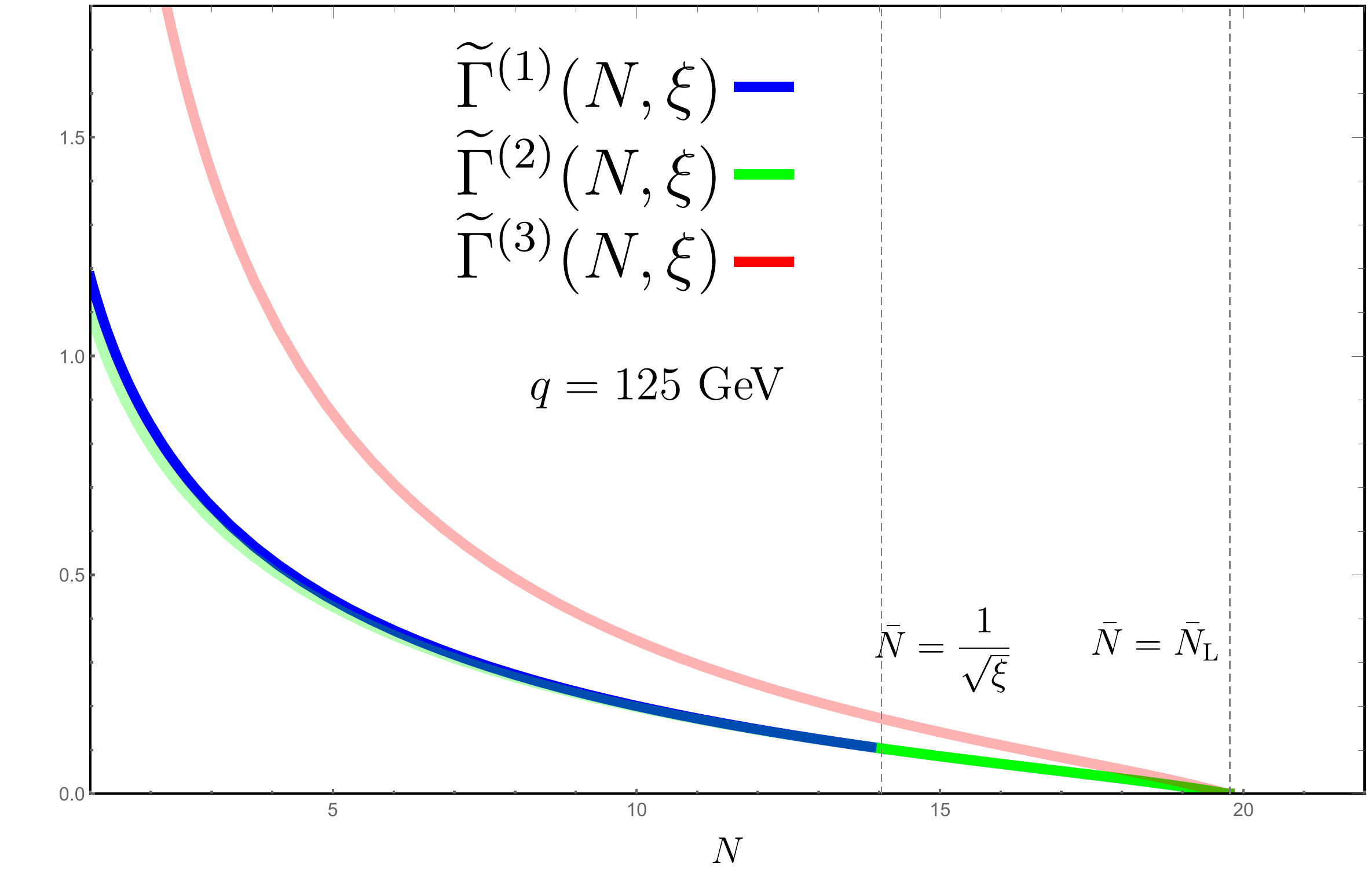}
	\caption{Our results, Eq.~(\ref{eq:NLL2}), plotted as a single function of $N$, on the real axis.
The colour-code for the different $\widetilde{\Gamma}^{(i)}$ is the same as in Fig.~\ref{fig:results-different-regions}, but now they are shown with solid lines within their $N$-space region of validity and transparent otherwise.} 
	\label{fig:results-one-region}
\end{figure}

It is also interesting to consider our Mellin space result Eq.~(\ref{eq:NLL2}) as a single function of $N$, on the real $N$ axis. This way, we can interpret the different $x$-space regions $\Theta$-functions in $N$-space, identifying $1-x \to \bar N^{-1}$.
Our results are shown in Fig.~\ref{fig:results-one-region}, again for $q=25$~GeV on the left, and $q=125$~GeV on the right. The colour-code for the different $\widetilde{\Gamma}^{(i)}$ is the same as before, but now they are shown with solid lines within their $N$-space region of validity and transparent otherwise. 
We note that, while the exponential functions in Eq.~(\ref{eq:NLL2}) are continuous across these regions, the subtracted contributions introduce discontinuities. However, because of the choice of the matching condition, Eq.~(\ref{eq:matching-choice}) the resulting function is actually continuous at $\bar N^{-1}=\sqrt{\xi}$ and the only discontinuity appears at $\bar N^{-1}=\xi$. This discontinuity has also been discussed in the literature, e.g.~\cite{Aglietti:2022rcm}.

We conclude this discussion by noting that our calculation can be generalised to the case of the decay of a massive colour singlet into two quarks with different masses, which is relevant, for instance, for the decay of a $W^{\pm}$ vector boson.
If we consider the emission of a gluon, we have
\begin{equation}
1-x = \frac{2 k \cdot p_2}{q^2}-\xi_1+\xi_2, \quad \xi_i=\frac{m_i^2}{q^2}.
\end{equation}
In order to work with an observable that vanishes on Born kinematics, we define a new scaling variable:
\begin{equation}\label{eq: chi}
	\chi=\frac{x}{1+\xi_1-\xi_2}.
\end{equation}
Clearly $\chi \to 1$ when the gluon becomes either soft or collinear to $p_2$. Thus, one should consider Mellin moments with respect to $\chi$
\begin{align}\label{eq:rate-frag-N-diff-mass}
\widetilde{\Gamma}(N,\xi_1,\xi_2)&= \frac{1}{\Gamma_0}\int_0^1 \de \chi \, \chi^{N-1}\,  \frac{\de\Gamma}{\de \chi}.
\end{align}
The calculation of the jet functions proceeds in the same way as described in the previous section, except that in this case the structure of integration regions is more complicated, due to the presence of two distinct thresholds.
We have not performed the explicit calculations, but we have checked that logarithms of $\xi_1$ are resummed by DGLAP evolution, while double logarithms of $\xi_2$ are exponentiated in the recoil jet function, as expected.

\section{Extension to other processes}\label{sec:extensions}

Thus far, we have focussed our discussion on the decay of a colour singlet into a heavy-quark pair. However, our calculation relies upon factorisation properties of QCD matrix elements in the quasi-collinear and soft limits, which are universal. Therefore, our formalism can be applied to different processes. In this section, we briefly discuss some applications, highlighting similarities and differences with other approaches. 
Detailed phenomenological studies at the level of physical processes are left to future work.

The first process we would like to discuss is heavy-quark decay
\begin{equation}
Q_1(q) \to Q_2(p_1)+V(p_2) + X(k),
\end{equation}
where $V$ is colour neutral boson, and $X$ stands for unresolved QCD radiation. 
A very interesting case for collider physics is $t \to  b W$.
In this context, one can combine a theoretical calculation that correctly treats the $b$-quark kinematics, with mass effects, together with a fragmentation function approach that is able to resum the large logarithms of $m_b/m_t$. 
In the limit where QCD radiation $X$ becomes either soft or collinear to the fermions' directions, all-order resummation becomes relevant. Soft-gluon resummation for top decay has been studied in \cite{Corcella:2001hz,Cacciari:2002re,Mitov:2003bm}. In these studies, one is usually interested in the kinematic distributions of the $b$-quark (or $B$ hadron). It follows that QCD evolution of the $b$-quark is always described by the measured jet function, and so there is no issue with non-commuting limits~\cite{Cacciari:2002re,Gaggero:2022hmv}. Therefore, the matching of the massless and massive soft-gluon resummed calculations is straightforward.

The situation is rather different for kinematic distributions of the vector boson decay products: the $W$ in the aforementioned $t\to  b W$ decay, or the photon in $b \to s \gamma$ processes. Soft gluon resummation in these contexts has been widely studied, e.g~\cite{Korchemsky:1994jb,Akhoury:1995fp,Leibovich:1999xfa,Leibovich:2000ig,Aglietti:2001cs,Aglietti:2001br,Aglietti:2001ng,Aglietti:2002ew,Aglietti:2022rcm,Becher:2005pd,Becher:2006qw,Galda:2022dhp}. In these examples, the final-state heavy quark is described by the recoil jet function and therefore the massless and soft limits no longer commute. 
This problem has been addressed by Aglietti~\emph{et al.}, who were able to derive a resummed expression that is valid in both massive and massless regimes~\cite{Aglietti:2007bp,Aglietti:2022rcm}. The basic idea is the same as the one we have described in this paper, namely one has to consider the quasi-collinear limit with respect to the unmeasured quark. However, in contrast to our results, Aglietti~\emph{et al.} find a smooth function (in momentum space) that interpolates between the two regimes, rather than a function which is defined by cases. 
In our understanding, the main difference between the two approaches is that, while we compute the unmeasured jet function using the massive splitting function, Aglietti~\emph{et al.} exploit the dipole formalism with massive partons~\cite{Catani:2002hc}. Because the massive dipole reduces to the splitting function in the quasi-collinear limit, the two approaches agree at NLL level, but treat non-logarithmic corrections (related to recoil against soft and collinear emissions) differently. 
It would be interesting to numerically compare the two approaches. However, it is not clear to us how flavour thresholds are implemented in Refs.~\cite{Aglietti:2007bp,Aglietti:2022rcm} and so we leave any detailed comparison to future work.

Another interesting process to be mentioned is deep-inelastic scattering (DIS) with heavy flavours. In particular, let us consider charged current (CC) DIS, either the scattering of a neutrino $\nu$ off a nucleus $N$, $\nu N \to l X$, where $l$ is a charged lepton, or $l N \to \nu X$ scattering. In both cases, we are interested in the situation in which charm flavour is identified in the final state. The parton-model contribution to these processes is given by
\begin{equation}
	q(p_1)+W^*(q)\to c (p_2)\quad p_1^2=0,\quad p_2^2=m_c^2,
\end{equation}
where the initial-state quark is taken massless.~\footnote{Neutral-current DIS with heavy flavours is also interesting, especially in the context  of studies about the intrinsic heavy-flavour component of the proton wavefunction.} 
It is interesting to note that CC-DIS experiments cover a rather large kinematical range in both the Bjorken scaling variable $x_\text{B}= \frac{Q^2}{2 P\cdot q}$, where $P$ is the momentum of the nucleon and  $Q^2=-q^2>0$ is the momentum transferred. 
Depending on the relative size of the charm quark mass $m_c^2$ and $Q^2$, a massless calculation or a massive one maybe more appropriate. 
Furthermore, if we want to investigate the large $x_{\text{B}}$ as done, for instance, in fixed-target experiments, then we face the issue of consistently combining the resummation of collinear logarithms and large-$x$ logarithms.
This problem was first investigated in Ref.~\cite{Corcella:2003ib}, where two separate formulations of soft-gluon resummation were derived: one valid in massless limit $m_c^2 \ll Q^2$ and one that kept the full charm-mass dependence, appropriate for the region $m_c^2 \sim Q^2$. 
The authors of Ref.~\cite{Corcella:2003ib} noted that the non-commutativity of the $m_c^2/Q^2 \to 0$ and $N \to \infty$ limits prevented them from obtaining a single resummed formula. 
Thanks to the formalism we have developed, we can bridge the gap between these two formulations, arriving at a unified resummed expression that interpolates between the two regimes.

Let us briefly describe the application of our formalism to this situation.
In DIS experiments, a measurement of $x_\text{B}$ probes the dynamics of the incoming particle.
Therefore, we can defined a measured jet function that describes radiation collinear to the initial-state quark. This is typically a massless jet function. However, note that, in contrast to the heavy-quark fragmentation case, where we had a perturbative initial condition, here we must supplement the description of radiation collinear to the initial state with non-perturbative PDFs, which also require specifying the flavour number scheme.
The unmeasured jet function instead describes radiation collinear to the final-state charm. Our calculation of $\bar j$ allows us to interpolate between the massless $m_c^2\ll Q^2$ and massive $m_c^2\sim Q^2$, thus consistently resumming both soft and mass logarithms.
Phenomenological studies concerning the application of our formalism to CC DIS are work in progress.

\section{Conclusions and outlook}\label{sec:conclusions}
In this paper, we have considered all-order calculations in processes with heavy flavours (mainly $b$ quarks) in the final state. We have first reviewed the two different calculational frameworks that are usually employed for these processes, namely the massive and the massless schemes. 
The former allows us to fully take into account, at a given perturbative order, the dependence on the quark mass, while the latter allows us to resum, to all perturbative orders, large logarithms of the ratio of the quark mass over the hard scale of the process. 
Massive and massless calculations are usually combined together to obtain more reliable theoretical predictions.
In particular, we have considered differential distributions, focussing on the energy fraction $x$ of a heavy quark, produced in $H \to b \bar b$ decays. The measurement of $x$ introduces a further scale and logarithms of $1-x$, which become large in the soft (or collinear) limit, should be resummed to all orders. We have reviewed the state-of-the-art results for soft-gluon resummation in both the massive and massless schemes. 

Ideally, one would like to combine massive and massless calculations, both supplemented with their own soft-gluon resummation. However, standard matching procedures do not work, essentially because of the non-commutativity of the $x\to 1$ and $m^2/q^2 \to 0$ limits. 
The main result of this paper is a modification of the massless calculation that makes this combination possible.
This is achieved by performing the calculation of the jet function related to the unmeasured $\bar b$ quark in the quasi-collinear limit, rather then in the strictly collinear one, as it usually done in the massless calculation. 
A second achievement of this work is a detailed analysis of the role played by the heavy-quark flavour threshold on resummed expressions. 

Our calculation is performed in momentum space and exploits the Lund representation in order to better identify the different kinematic regions. It is then transformed to Mellin space to next-to-leading logarithmic accuracy, so that it can be easily combined with results present in the literature.
Thanks to our results, we can consistently resum soft and mass logarithms, while taking into account the effect of heavy-quark thresholds. 
While we leave detailed phenomenology to future work, we have performed some studies in Mellin space. We find that the effects we are studying become relevant only for rather low invariant masses, while are small for on-shell Higgs decay. 
Nevertheless, it would be interesting to combine our formalism with recent high-precision studies of heavy-quark fragmentation, see e.g.~\cite{Maltoni:2022bpy,Czakon:2022pyz,Bonino:2023vuz}, and investigate our approach beyond NLL, where because of soft-gluon interference effects, it is no longer possible to factorise the resummation formula into the product of two independent jet functions.

We have also discussed other processes that present similar features, namely heavy-quark decay, commenting on similarities and differences of our approach and the one by Aglietti~\emph{et al.}, and DIS with heavy flavours.
We find the DIS case particularly interesting because experimental data span several orders of magnitude in $Q^2$. We plan to focus on these processes for future phenomenological studies. 
%

We conclude by noting that, as a by-product of our study, we have reached a detailed understanding of the use of the Lund plane to perform resummed calculations with massive quarks. In conjunction with recently developed IRC safe flavour-jet algorithms~\cite{Caletti:2022hnc,Czakon:2022wam,Gauld:2022lem,Caola:2023wpj}, this progress will allow us to extend existing resummed calculations for jet substructure to the case of heavy-flavour jets. These include, for instance, jet angularities~\cite{Caletti:2021oor,Reichelt:2021svh,Kang:2018vgn}, Soft Drop observables~\cite{Dasgupta:2013ihk,Larkoski:2014wba,Larkoski:2015lea,Kang:2019prh,Cal:2021fla} and the primary Lund plane density~\cite{Lifson:2020gua}, opening up exciting possibility to perform precision heavy-flavoured jet substructure studies at the LHC. 

\acknowledgments

We thank Matteo Cacciari, Fabrizio Caola, Giancarlo Ferrera, Gavin Salam, Gregory Soyez, and Maria Ubiali for useful discussions on this topic. 
AG and SM acknowledge support from the IPPP DIVA fellowship program and thank the physics department at the University of Oxford for hospitality during the course of this work. 
AG and SM also thank the Erwin-Schr\"odinger International Institute for Mathematics and Physics at the University of Vienna for partial support during the Programme ``Quantum Field Theory at the Frontiers of the Strong Interactions", July 31 - September 1, 2023.
The work of GR is supported by the Italian Ministero dell’Universit\`a e della Ricerca (MUR) under grant PRIN 20172LNEEZ.

\appendix

\section{Resummed expressions in the massless scheme}\label{Resummation:FF}
In this Appendix we report, for completeness, explicit results for  the resummed differential rate in the fragmentation-function approach, Eq. (\ref{fragm-master}).
The two-loop running coupling with $n$ active flavours is given by
\begin{equation}
	\as(k_t^2)=\frac{\as(\mu^2)}{1+\nu}\left(1-\frac{\beta_1^{(n)}}{\beta_0^{(n)}}\as(\mu^2)\frac{\log{\left(1+\nu\right)}}{1+\nu}\right),\quad \nu=\as(\mu^2)\beta^{(n)}_0\log{\frac{k_t^2}{\mu^2}}
\end{equation}
with
\begin{align}
	\beta_0^{(n)}=\frac{11\ca-2 n}{12\pi}\qquad \beta_1^{(n)}=\frac{17\ca^2-5\ca n-3\cf n}{24\pi^2}
\end{align}
and $\ca=3,\cf=\frac{4}{3}$.

Non-singlet DGLAP evolution at next-to-leading logarithmic accuracy is described by the following evolution kernel (see \cite{Mele:1990yq}):
\begin{align}\label{eq:dglap-evolution_Kernel} 
	\log \mathcal{E}(N,\muOf^2,\muf^2)&=
	\frac{\gamma^{(0)}(N)}{\pi \beta_0^{(n_f)}}
	\log\frac{\as(\muOf^2)}{\as(\muf^2)}\nonumber\\
	&+
	\frac{\as(\muOf^2)-\as(\muf^2)}{\pi^2 \beta_0^{(n_f)}}
	\left[\gamma^{(1)}(N)-\frac{\pi \beta_1^{(n_f)}}{\beta_0^{(n_f)}}\gamma^{(0)}(N)\right],
\end{align}
where we have introduced the DGLAP anomalous dimension:
\begin{equation}
	\gamma(N,\as)= \frac{\as}{\pi} \gamma^{(0)}(N)+\left(\frac{\as}{\pi}\right)^2 \gamma^{(1)}(N) +\order{\as^3}.
\end{equation}
We have
\begin{equation}
	\gamma^{(0)}(N)=-\frac{\cf}{2} \left[2 \left(\psi_0(N)+\gamma_E\right)-\frac{3}{2}+\frac{1}{N}+\frac{1}{N+1} \right],
\end{equation}
where 
\begin{equation}
	\psi_0(x)=\frac{\de}{\de x} \log{\Gamma(x)}.
\end{equation}
An explicit expression for $\gamma^{(1)}(N)$ can be found for example in \cite{Mele:1990cw}.

The resummed initial condition for the fragmentation function was computed in \cite{Cacciari:2001cw}:
\begin{align}\label{eq:resum-initial}
	 \widetilde{\mathcal{D}}_0\left(N,\frac{\muOf^2}{m^2},\frac{\muOr^2}{m^2}\right)&=\left(1+ \frac{\as \cf}{\pi}\mathcal{D}_0^{(1)} \right)\exp D_0 \left(N, \frac{\muOf^2}{m^2},\frac{\muOr^2}{m^2} \right),
\end{align}
with
\begin{align} \label{eq:resum-initial-detail}
	D_0&\left(N, \frac{\muOf^2}{m^2},\frac{\muOr^2}{m^2} \right)=d_1\log \bar{N} +d_2,\\ \nonumber
	d_1&=-\frac{A_1}{2 \pi \beta_0^{(n_l)} \lambda_0}\left[ 2\lambda_0+ (1-2\lambda_0) \log(1-2\lambda_0)\right], \nonumber\\   
	d_2&= -\frac{A_1}{2 \pi}\frac{\beta_1^{(n_l)}}{\beta_0^{(n_l)3}}\left[2\lambda_0+\log(1-2\lambda_0)+\frac{1}{2}\log^2(1-2\lambda_0) \right]\nonumber\\
	&+\left(\frac{A_2^{(n_l)}}{2 \pi^2 \beta_0^{(n_l)2}}+\frac{A_1}{2\pi \beta_0^{(n_l)}} \log{\frac{\muOr^2}{\muOf^2}}\right) \left[2\lambda_0+\log(1-2\lambda_0) \right]+\frac{H_1}{2 \pi \beta_0^{(n_l)}}\log(1-2 \lambda_0)
	\nonumber\\
	&+\frac{A_1}{2 \pi \beta_0^{(n_l)}}\log{(1-2\lambda_0)}\log \frac{\muOf^2}{m^2},
	\nonumber\\   
	\mathcal{D}_0^{(1)}&= 1-\frac{\pi^2}{6}+\frac{3}{4}\log\frac{\muOr^2}{m^2},
\end{align}
and we have defined $\lambda_0=\as^{(n_l)}(\muOf^2)\beta_0^{(n_l)}\log{\bar{N}}$, $n_l=4$.

The resummed coefficient function is given by:
\begin{align}\label{eq:resum-massless}
\mathcal{C}\left(N,\frac{\muf^2}{q^2},\frac{\mur^2}{q^2},\as(\mur^2)\right)&=\left(1+ \frac{\as \cf}{\pi}\mathcal{C}^{(1)} \right)\exp \left[ \Delta \left(N,\frac{\muf^2}{q^2},\frac{\mur^2}{q^2}\right)+ \bar J\left(N,\frac{\mur^2}{q^2}\right)\right] ,
\end{align}
where both $\Delta$ and $\bar J$ are computed in the strict collinear (massless) limit:
\begin{align}\label{eq:resum-massless detail}
	\Delta&\left(N,\frac{\muf^2}{q^2},\frac{\mur^2}{q^2}\right)+ \bar J\left(N,\frac{\mur^2}{q^2}\right)= c_1\log{\Bar{N}} +c_2,\\
	\notag c_1=&\frac{A_1}{\pi\beta_0^{(n_f)} }\frac{\left(\lambda+(1-\lambda)\log{(1-\lambda)}\right)}{\lambda},\\
	\notag c_2=&
	\left(\frac{A_1}{\pi\beta_0^{(n_f)}}\log\frac{q^2}{\mur^2}-\frac{A_2^{(n_f)}}{\pi^2\beta_0^{(n_f)2}}\right)\Big(\lambda+\log{(1-\lambda)}\Big)-\frac{A_1}{\pi \beta_0^{(n_f)}}\lambda \log \frac{q^2}{\muf^2}\\ \notag
	&+\frac{A_1\beta_1^{(n_f)}}{2\pi\beta_0^{(n_f)3}}\Big(\log^2{(1-\lambda)}+2\log{(1-\lambda)}+2\lambda\Big)
+\frac{B_1}{2\pi \beta_0^{(n_f)}}\log{(1-\lambda)},
\end{align}
and $\lambda=\as(\mur^2)\beta_0^{(n_f)}\log{\bar{N}}$, $n_f=5$.
The only difference between photon, Higgs and $Z$ boson is due to the overall constant $\mathcal{C}^{(1)}$ (see \cite{Corcella:2004xv} and \cite{Mele:1990cw}):
\begin{align} \label{C different processes}
	\text{for} \;H\to b\bar{b}:& \qquad \mathcal{C}^{(1)}= \frac{3}{4}+\frac{5}{12}\pi^2+\frac{3}{4}\log \frac{q^2}{\muf^2},\\
	\text{for} \;Z/\gamma^*\to b\bar{b}:&\qquad \mathcal{C}^{(1)}= -\frac{9}{4}+\frac{5}{12}\pi^2+\frac{3}{4}\log \frac{q^2}{\muf^2}.
\end{align}

\section{Mellin moments to next-to-leading logarithmic accuracy}
\label{sec:moments}

In this Appendix we show that the Mellin transform of momentum-space resummed results, such as those obtained in Sect.~\ref{sec:jet_mom}, can be computed, to next-to-leading logarithmic accuracy, by the replacement
\begin{equation}
\log(1-x)\to \log\frac{1}{\bar N},
\end{equation}
where 
\begin{equation}
\bar N=Ne^{\gamma_{\rm E}}
\end{equation}
and $N$ is the Mellin variable conjugate to $x$.

We consider specifically the normalized cumulative energy distribution
\begin{equation}
\Sigma(x,\xi)=\int_x^1\de x' \, \Gamma(x',\xi);\qquad \Gamma(x,\xi)=\frac{1}{\Gamma_0} \frac{\de \Gamma}{\de x'}.
\end{equation}
Its Mellin transform can be related to the Mellin transform $\widetilde{\Gamma}(N,\xi)$ of the normalised distribution. We find
\begin{equation}
\widetilde{\Sigma}(N,\xi)=\int_0^1dx\,x^{N-1}\int_x^1\de x' \, \Gamma(x',\xi)
=\frac{1}{N}\widetilde{\Gamma}(N+1,\xi),
\end{equation}
or
\begin{equation}
N\widetilde{\Sigma}(N,\xi)=\widetilde{\Gamma}(N+1,\xi).
\label{sigmatilde0}
\end{equation}
On the other hand, the perturbative expansion of $\Sigma(x,\xi)$ at large $x$ has the form
\begin{equation}
\Sigma(x,\xi)= \sum_{k=0}^{+\infty}\as^k\sum_{n=0}^{2k} c_n(\xi) \log^{2k-n}(1-x),
\end{equation}
where the coefficients $c_n(\xi)$ are polynomials in $\log\xi$. We have
\begin{align}
\int_0^1dx\,x^{N-1}\log^p (1-x)&= \frac{\de^p }{\de a^p}\int_0^1dx\,x^{N-1} (1-x)^a \Big|_{a=0}
\nonumber\\
&=\frac{\de^p }{\de a^p}\frac{\Gamma(a+1)\Gamma(N)}{\Gamma(a+1+N)}\Big|_{a=0}
\nonumber\\
&=\frac{\de^p }{\de a^p}\frac{\Gamma(a+1)}{a+N}\left(\frac{1}{N^a}+ \mathcal{O}\left( N^{-1} \right)\right)\Big|_{a=0}.
\end{align}
To NLL  accuracy we have therefore
\begin{align}
N\int_0^1dx\,x^{N-1}\log^p (1-x)&=\log^p\frac{1}{N}-p\gamma_E\log^{p-1}\frac{1}{N}+\mathcal{O}\left(\log^{p-2}\frac{1}{N}\right)
\nonumber\\
&=\log^p\frac{1}{\bar N}+\mathcal{O}\left(\log^{p-2}\frac{1}{N}\right).
\end{align}
This identity allows us to obtain a NLL expression for the Mellin transform of the normalised cumulative distribution:
\begin{equation}
N\widetilde{\Sigma}(N,\xi)=\left. N\int_0^1 \de x \, x^{N-1} \Sigma(x,\xi)=\Sigma\left(x,\xi\right)\right|_{1-x=\frac{1}{\bar N}}
\label{sigmatilde}
\end{equation}
Comparing Eqs. (\ref{sigmatilde0}) and~(\ref{sigmatilde}) we finally obtain
\begin{equation}
\widetilde{\Gamma}(N+1,\xi)=\Sigma\left(x,\xi\right)\Big|_{1-x=\frac{1}{\bar N}}
\end{equation}
up to NNLL corrections, which is the announced result.

The case we are interested in presents a further subtlety, because we want to compute moments of a function that is defined by cases:
\begin{align}\label{eq:Mellin vs Comulative}
	\Sigma(x,\xi)&=
	 \exp\Bigg\{\left [ j^{(1)}(1-x,\xi)+ \bar j^{(1)}(1-x,\xi) \right]\Theta(1-x -\sqrt{\xi})
	\nonumber \\ &+\exp \left [ j^{(2)}(1-x,\xi)+ \bar j^{(2)}(1-x,\xi) \right] \Theta(1-x -\xi)\Theta(\sqrt{\xi}-1+x)\nonumber \\ &+ \exp \left [ j^{(2)}(1-x,\xi)+ \bar j^{(3)}(1-x,\xi) \right]\Theta(\xi-1+x)\Bigg\}.
\end{align}
However
\begin{align}
& N \int_0^{1-\sqrt{\xi}} \de x \, x^{N-1} (1-x)^a =\frac{\Gamma(1+a)}{ N^a}+ \mathcal{O}\left( N^{-1},\sqrt{\xi} \right),   \nonumber \\
 & N \int^{1-\xi}_{1-\sqrt{\xi}} \de x \, x^{N-1} (1-x)^a =\frac{\Gamma(1+a)}{ N^a}+ \mathcal{O}\left( N^{-1}, \xi, 1-\sqrt{\xi} \right),   \nonumber \\
 & N \int_{1-\xi}^1 \de x \, x^{N-1} (1-x)^a = \frac{\Gamma(1+a)}{ N^a}+ \mathcal{O}\left( N^{-1}, 1-\xi\right).
  \end{align}
We conclude that the $\Theta$ functions that establish the boundaries between different regions of the measured and unmeasured jet functions do not alter, in each region, the correspondence between logarithms of $1-x$ and logarithms of $\bar N$ nor introduce logarithms of $\xi$. Thus, in each region, Mellin moments of $j^{(i)}$ and $\bar j^{(i)}$ in Eq.~(\ref{eq:jet-function-thetas}), and~(\ref{eq:recoil-jet-function-thetas})  can be found by simply replacing $\log(1-x) \to - \log \bar N$.

\section{Details of the running coupling integrals}\label{app:running_coupling}
In this section we outline the technique employed to compute the integrals with running coupling in the region where mass effects are not negligible. 
In particular we focus on the computation of Eq.~(\ref{eq:jet-function-Jbar}).  The same technique applies to Eq.~(\ref{eq:jet-function-J}). Eq.~(\ref{eq:jet-function-Jbar}) can be written
\begin{equation}\label{eq:barJ changed variables}
	\bar j(1-x,\xi)= \int^{1}_{\text{max}{\left(\xi,1-x\right)}} \frac{\de \zeta}{\zeta}\int^1_{\frac{1-x}{\zeta}} \de  z_2 \; \frac{\as^{\text{CMW}}(z_2^2(\zeta-\xi)q^2) }{2 \pi}P_{gq}( z_2,  z_2^2(\zeta-\xi)q^2),
\end{equation}
where we have changed one integration variable to $\zeta= \frac{k_t^2+z_2^2 m^2}{q^2  z_2^2}$. The strong coupling may now be expressed as a power series in $\log{( z_2^2(\zeta-\xi))}$.
 We focus on the most singular term in the splitting function, and so we consider the integral:
\begin{equation}
	I_p=	\as^{p+1} \int^1_{\max{\left(\xi,1-x\right)}} \frac{\de \zeta}{\zeta} \int^1_{\frac{1-x}{\zeta}}\frac{\de  z_2}{z_2} \log^p{\left(z_2^2(\zeta-\xi)\right)}. 
\end{equation}
Performing the integration over $z_2$, we obtain
\begin{equation}
	I_p=-\as^{p+1} \sum_{k=0}^p \binom{p}{k} \frac{2^{p-k}}{p-k+1} \int^1_{\max{\left(\xi,1-x\right)}} \frac{\de \zeta}{\zeta} \log^k{\left(\zeta-\xi\right)}\log^{p-k+1}{\left(\frac{1-x}{\zeta}\right)}
\end{equation}
In order to perform the $\zeta$ integration, we consider the Taylor series of the integrand with respect to $\xi$ about $\xi=0$:
\begin{equation}\label{eq:first step}
	\begin{split}
			I_p=&-\as^{p+1} \sum_{k=0}^p \binom{p}{k} \frac{2^{p-k}}{p-k+1} \sum_{l=0}^\infty \frac{\xi^l}{l!}\\
			&\int^1_{\max{\left(\xi,1-x\right)}} \frac{\de \zeta}{\zeta} \frac{\de^l}{\de^l \xi}\log^k{\left(\zeta-\xi\right)}|_{\xi=0}\log^{p-k+1}{\left(\frac{1-x}{\zeta}\right)}.
	\end{split}
\end{equation}

Let us consider first the situation $1-x>\xi$. In this case, up to power corrections in $\xi$, we can neglect the $\xi$ dependence in the integrand and keep only the $l=0$ contribution in the sum. This way, we recover the standard (massless) result. 

The second case, $1-x<\xi$, requires a more careful treatment.
We notice that
\begin{equation} \label{eq:derivative expansion}
	\frac{\de^l }{\de^l \xi} \log^k{\left(\zeta-\xi\right)}|_{\xi=0}= \delta_{l,0}\log^k{\zeta}+(1-\delta_{k,0})(1-\delta_{l,0})\frac{1}{\zeta^l} \sum^{k-1}_{n=0} a_n \log^n{\zeta},
\end{equation}
with $a_n=a_n(k,l)$. The previous relation can be proved by induction. Inserting Eq.~(\ref{eq:derivative expansion}) in Eq.~(\ref{eq:first step}),
\begin{equation} \label{eq:second step}
	\begin{split}
		I_p=&-\as^{p+1} \sum^p_{k=0}\binom{p}{k} \frac{2^{p-k}}{p-k+1}\sum^\infty_{l=0} \frac{\xi^l}{l!} 
		\\& \int_{\xi}^1 \frac{\de \zeta}{\zeta} \left[\delta_{l,0}\log^k{\zeta}+(1-\delta_{k,0})(1-\delta_{l,0})\frac{1}{\zeta^l} \sum^{k-1}_{n=0} a_n \log^n{\zeta} \right]  \log^{p-k+1}\left(\frac{1-x}{\zeta}\right),		
		\end{split}
\end{equation}
 and performing the integral over $\zeta$ we find:
\begin{equation} \label{eq:third step}
	\begin{split}
		I_p=&\as^{p+1} \sum^p_{k=0}\binom{p}{k} \frac{2^{p-k}}{p-k+1}\sum^\infty_{l=0} \frac{1}{l!} \sum_{j=0}^{p-k+1}(-1)^j 
		\binom{p-k+1}{j}\log^{p-k+1-j}(1-x)\\
		&\left(\delta_{l,0}\frac{\log^{k+j+1}{\xi}}{k+j+1}-(1-\delta_{k,0})(1-\delta_{l,0})
		\frac{a_{k-1} \log^{k-1+j}{\xi}}{l}\right)+\order{\as\left(\as \log{\xi}\right)^p}.
		\end{split}
\end{equation}
The generic term of the above series has the form $\as^{p+1} \log^a(1-x) \log^b{\xi}$.
The leading contribution is the one with $a+b=p+2$ and is given entirely by the $l=0$ contribution, i.e.\ the first term in parenthesis of the second line of Eq.~(\ref{eq:second step}).
The first correction arising from the second term in the sum instead gives at most $a+b=p$. Therefore, we conclude that the mass shift in the running coupling gives contributions only beyond NLL and, consequently, it can be safely dropped in our calculations. 

\section{Resummed expressions for the jet functions}\label{app:calculations}
In this appendix we provide explicit results of the resummed expression for the measured and unmeasured jet functions in Mellin space. We write our results for generic $n_l$ and $n_f=n_l+1$ active flavours, even if in this work we have only explicitly considered $n_l=4$ and $n_f=5$.

\subsection{Resummed expressions for the measured jet function}
We present the explicit results for the measured jet functions $\tilde{j}^{(i)}$, $i=1,2,3$.

In the first region for $1-x<\sqrt{\xi}$, $\tilde{j}^{(1)}=J=D_0+E+\Delta$, see Eq.~(\ref{eq:j-def}). The first contribution is computed with $n_l$ flavours and is simply given by Eq.~(\ref{eq:resum-initial-detail}). 
The second contribution, with $n_f$ flavours, obtained by computing the integral in Eq.~(\ref{eq:evolution operator integral}):
\footnote{This is not strictly true because the factorisation scale $\muOf$ is not necessarily equal to $m$. So, if for instance $\muOf<m$, DGLAP evolution also contains a (small) $n_l$-flavour contribution. On the other hand, if $\muOf>m$, the initial condition $D_0$ has a $n_f$-flavour piece. Because $\muOf$ is always chosen of the order of $m$, where are going to ignore such complications.}
\begin{equation}
	\begin{split} \label{eq:explicit E}
E(N,\muOf^2,\muf^2)&=\log \bar N \Bigg[
\frac{A_1}{\pi \beta_0^{(n_f)}}\log\frac{\as(\muf^2)}{\as(\muOf^2)}+
\left(\as(\muf^2)-\as(\muOf^2)\right)
\left(\frac{A^{(n_f)}_2}{\pi^2\beta_0^{(n_f)}}-\frac{A_1 \beta_1^{(n_f)}}{\pi\beta_0^{(n_f)2}}\right)\Bigg]
\\
 &+\frac{B_1}{2\pi \beta_0^{(n_f)}} \log\frac{\as(\muf^2)}{\as(\muOf^2)}.
\end{split}
\end{equation}
The third term is obtained from Eq.~(\ref{eq: delta function}) and reads
\begin{equation}
	\begin{split}
		\Delta^{(1)}&\left(N,\frac{\muf^2}{q^2},\frac{\mur^2}{q^2}\right)= \log{\Bar{N}} \delta_1^{(1)}+\delta_2^{(1)},\\
		 \delta_1^{(1)}&=\frac{A_1}{2\pi\beta^{(n_f)}_0 \lambda}\left(2\lambda+(1-2\lambda)\log{(1-2\lambda)}\right),\\
		\notag \delta_2^{(1)}&= \frac{A_1}{2\pi\beta_0^{(n_f)}}\left[-2\lambda\log{\left(\frac{q^2}{\muf^2}\right)}+\log{\left(\frac{q^2}{\mur^2}\right)}\Big(2\lambda+\log{(1-2\lambda)}\Big)\right]\\
		&\notag+\frac{A_1\beta_1^{(n_f)}}{4\pi\beta_0^{(n_f)3}}\Big(\log^2{(1-2\lambda)}+2\log{(1-2\lambda)}+4\lambda\Big)\\
		\notag&-\frac{A_2^{(n_f)}}{2\pi^2\beta_0^{(n_f)2}}\Big(2\lambda+\log{(1-2\lambda)}\Big),
	\end{split}
\end{equation}
The suffix $^{(1)}$ indicates that we are in the first region, $1-x>\sqrt{\xi}$.

In the second region $1-x< \sqrt{\xi}$ the jet function $\tilde{j}^{(2)}$ is obtained by computing the integrals in Eq.~(\ref{eq:J-region2}). In particular, we have
\begin{align} \label{eq:explicit J2}
\tilde{j}^{(2)}(N,\xi)&= D_0\left(N,\frac{\muOf^2}{m^2},\frac{\muOr^2}{m^2}\right)+ E(N,\muOf^2,\muf^2)+\hat E\left(N,\xi,m^2,\muf^2\right)\nonumber \\&+\Delta^{(1)}\left(\frac{e^{-\gamma_E}}{\sqrt{\xi}},\frac{\muf^2}{q^2},\frac{\mur^2}{q^2} \right)+\Delta^{(2)}\left( N,\xi,\frac{\muOr^2}{m^2}\right),
\end{align}
with     
\begin{equation}
	\begin{split}
\hat E \left(N,\xi,m^2,\muf^2\right)=& \int^{\sqrt{\xi}}_{\frac{1}{\bar N}} \de z P_{gq}(z)\int^{\muf^2}_{m^2}\frac{\de k_t^2}{k_t^2} \frac{\as^{\text{CMW}}(k_t^2)}{\pi}
\\ -&\log{\left(\bar N\sqrt{\xi}\right)} \Bigg[
\frac{A_1}{\pi \beta_0^{(n_f)}}\log\frac{\as(\muf^2)}{\as(m^2)}+
\left(\as(\muf^2)-\as(m^2)\right)
\left(\frac{A^{(n_f)}_2}{\pi^2\beta_0^{(n_f)}}-\frac{A_1 \beta_1^{(n_f)}}{\pi\beta_0^{(n_f)2}}\right)\Bigg].
\end{split}
\end{equation}  
We notice that, setting $\muOf^2=m^2$ in Eq.~(\ref{eq:explicit E}), $\tilde{j}^{(2)}$ depends on $N$ only through the $n_l$ contribution.
 The function $\Delta^{(2)}$ in Eq.~(\ref{eq:explicit J2}) is given by
\begin{equation}
\begin{split} \label{eq:delta2}
	\Delta^{(2)}\left(N,\xi,\frac{\muOr^2}{m^2}\right)=&\int^{\sqrt{\xi}}_{\frac{1}{\bar N}} \de z \int^{m^2}_{q^2 z^2} \frac{\de k_t^2}{k_t^2} \frac{\as^{\text{CMW}}(k_t^2)}{2\pi}P_{gb}(z,k_t^2-z^2m^2)\\
	=&\log{\bar N} \delta_1^{(2)}+\delta_2^{(2)}.
\end{split}
\end{equation}
where:
\begin{equation}
	\begin{split} \label{eq: delta_i 1 and 2}
			\delta_1^{(2)}&=\frac{A_1}{2 \pi \beta_0^{(n_l)} \lambda_0}\left[2\lambda_0-\lambda_{0\xi} +(1-2\lambda_0+\lambda_{0\xi}) \log(1-2\lambda_0+\lambda_{0\xi})\right],\\   
		\delta_2^{(2)}&= \frac{A_1}{2 \pi}\frac{\beta_1^{(n_l)}}{\beta_0^{(n_l)3}}\left[2\lambda_0-\lambda_{0\xi}+\log(1-2\lambda_0+\lambda_{0\xi})+\frac{1}{2}\log^2(1-2\lambda_0+\lambda_{0\xi}) \right]\\
		&-\frac{A_2^{(n_l)}}{2 \pi^2 \beta_0^{(n_l)2}} \left[2\lambda_0-\lambda_{0\xi}+\log(1-2\lambda_0+\lambda_{0\xi}) \right]
	\\
		&-\frac{A_1}{2 \pi \beta_0^{(n_l)}}\left[2\lambda_0-\lambda_{0\xi}+\log{(1-2\lambda_0+\lambda_{0\xi})}\right]\log \frac{\muOr^2}{m^2}.
	\end{split}
\end{equation}
Here we have defined $\lambda_{0\xi}=\as^{(n_l)}(\muOr^2) \beta_0^{(n_l)}\log{\frac{1}{\xi}}$.  We note that in the limit $\lambda_{0\xi}\ll1$ only the logarithmic term with coefficient $H_1$ survives.

\subsection{Resummed expressions for the unmeasured jet function}
We present the explicit results for the unmeasured (recoil) jet functions $\tilde{\bar{j}}^{(i)}$, $i=1,2,3$. We start by considering the region $1-x> \sqrt{\xi}$ and we integrate Eq.~(\ref{eq:Jbar_FF}). Upon the usual identification $1-x=\bar N^{-1}$, we find
\begin{align}\label{eq:recoil-jet}
\tilde{\bar{j}}^{(1)}&\left(N,\frac{\mur^2}{q^2}\right)=\bar g_1^{(1)}\log{\Bar{N}} + \bar g_2^{(1)},\\
    \notag \bar g_1^{(1)}&=\frac{A_1}{2\pi\beta_0^{(n_f)}}\frac{2(1-\lambda)\log{\left(1-\lambda\right)}-(1-2\lambda)\log{\left(1-2\lambda\right)}}{\lambda},\\
    \notag \bar g_2^{(1)}&=\frac{A_1}{2\pi\beta_0^{(n_f)}}\log\frac{q^2}{\mur^2}\Big(2\log{(1-\lambda)}-\log{(1-2\lambda)}\Big)\\
&\notag+\frac{A_1\beta_1^{(n_f)}}{4\pi{\beta_0^{(n_f)3}}}\Big[2\log{(1-\lambda)}\big(2+\log{(1-\lambda)}\big)-\log{(1-2\lambda)}\big(2+\log{(1-2\lambda)}\big)\Big]\\
&\notag-\frac{A_2^{(n_f)}}{2\pi^2\beta_0^{(n_f)2}}\Big(2\log{(1-\lambda)}-\log{(1-2\lambda)}\Big)+\frac{B_1}{2\pi\beta_0^{(n_f)}}\log{(1-\lambda)}.
\end{align}
The unmeasured jet function to be used in the region $\xi< 1-x<\sqrt{\xi}$ is then obtained by computing Eq~(\ref{eq:Jbar_intermediate_region}) and it is given by:
\begin{equation}
	\begin{split} \label{eq: Jbar2}
\tilde{\bar{j}}^{(2)}&\left(N,\xi,\frac{\mur^2}{q^2},\frac{\muOr^2}{m^2}\right)=\log{\bar{N}} \bar{g}_{1}^{(2)}+\bar{g}_{2}^{(2)} ,\\
		\bar{g}_{1}^{(2)}=&\frac{A_{1}}{2\pi\beta_0^{(n_l)}} \frac{-2\lambda_0+\lambda_{0\xi}-(1-2\lambda_0+\lambda_{0\xi})\log{\left(1-2\lambda_0+\lambda_{0\xi}\right)}}{\lambda_0}\\
		+	& \frac{A_{1}}{2\pi\beta_0^{(n_f)}}\frac{2\lambda-\lambda_\xi+2(1-\lambda)\log{(1-\lambda)}-(1-2\lambda)\log{(1-\lambda_\xi)}}{\lambda},\\
		\bar{g}_{2}^{(2)}=& -\frac{A_{1}}{2\pi\beta_0^{(n_l)}}\log\frac{m^2}{\muOr^2}\left(2\lambda_0-\lambda_{0\xi}+\log{\left(1-2\lambda_0+\lambda_{0\xi}\right)}\right)\\
		+&\frac{A_{1}}{2\pi\beta_0^{(n_f)}}\log\frac{q^2}{\mur^2}\frac{2\lambda-\lambda_\xi+(1-\lambda_\xi)(2\log{(1-\lambda)}-\log{(1-\lambda_\xi)})}{1-\lambda_\xi}\\
		+&\frac{A_2^{(n_l)}}{2\pi^2\beta_0^{(n_l)2}}\left(2\lambda_0-\lambda_{0\xi}+\log{\left(1-2\lambda_0+\lambda_{0\xi}\right)}\right)-\frac{A_2^{(n_f)}}{2\pi^2\beta_0^{(n_f)2}}\Big(\frac{2\lambda-\lambda_\xi}{1-\lambda_\xi}+2\log{(1-\lambda)}-\log{(1-\lambda_\xi)}\Big)\\
		-&\frac{A_{1}\beta_1^{(n_l)}}{4\pi\beta_0^{(n_l)3}}\left(4\lambda_0-2\lambda_{0\xi}+\log{(1-2\lambda_0+\lambda_{0\xi})}\left(2+\log{(1-2\lambda_0+\lambda_{0\xi})}\right)\right)\\
		+&\frac{A_{1}\beta_1^{(n_f)}}{4\pi\beta_0^{(n_f)3}}\Big\{ 2\Big[\frac{2\lambda-\lambda_\xi}{1-\lambda_\xi}+\log{(1-\lambda)}(2+\log{(1-\lambda)})\Big]-\log{(1-\lambda_\xi)}\left(2\frac{1-2\lambda}{1-\lambda_\xi}+\log{(1-\lambda_\xi)}\right)\Big\}\\
		+& \frac{B_{1}}{2\pi \beta_0^{(n_f)}}\log{(1-\lambda)}.
	\end{split}
\end{equation} 
where $\lambda_\xi=\as^{(n_f)}(\mur) \beta_0^{(n_f)}\log{\frac{1}{\xi}}$. Finally, in the third region, $1-x<\xi$, Eq~(\ref{eq:Jbar-final-region}) gives:
\begin{equation} \label{eq: Jbar3}
	\begin{split}
		\tilde{\bar{j}}^{(3)}&\left(N,\xi,\frac{\mur^2}{q^2},\frac{\muOr^2}{m^2}\right)=\log{\bar{N}} \bar{g}_{1}^{(3)}+\bar{g}_{2}^{(3)},\\
		\bar{g}_1^{(3)}=& \frac{A_{1}}{2\pi\beta_0^{(n_l)}} \frac{(1-2\lambda_0+2\lambda_{0\xi})\log{(1-2\lambda_0+2\lambda_{0\xi})}-\lambda_{0\xi}-(1-2\lambda_0+\lambda_{0\xi})\log{(1-2\lambda_0+\lambda_{0\xi})}}{\lambda_0}\\
		+& \frac{A_{1}}{2\pi\beta_0^{(n_f)}}\frac{\lambda_\xi+\log{(1-\lambda_\xi)}}{\lambda},\\
		\bar{g}_2^{(3)}=& \frac{A_{1}}{2\pi\beta_0^{(n_l)}}\log\frac{m^2}{\muOr^2}\Big(\log{(1-2\lambda_0+2\lambda_{0\xi})}-\log{(1-2\lambda_0+\lambda_{0\xi})}-\lambda_{0\xi}\Big)\\
		+&\frac{A_{1}}{2\pi\beta_0^{(n_f)}}\log\frac{q^2}{\mur^2}\frac{\lambda_\xi+(1-\lambda_\xi)\log{(1-\lambda_\xi)}}{1-\lambda_\xi}\\
		+& \frac{A_2^{(n_l)}}{2\pi^2\beta_0^{(n_l)2}}\left(\lambda_{0\xi}+\log{\left(\frac{1-2\lambda_0+\lambda_{0\xi}}{1-2\lambda_0+2\lambda_{0\xi}}\right)}\right)
		-\frac{A_2^{(n_f)}}{2\pi^2\beta_0^{(n_f)2}}\Big(\frac{\lambda_\xi}{1-\lambda_\xi}+\log{(1-\lambda_\xi)}\Big)\\
		-&\frac{A_{1}\beta_1^{(n_l)}}{4\pi\beta_0^{(n_l)3}}\left(2\lambda_{0\xi}+\log{\left(\frac{1-2\lambda_0+\lambda_{0\xi}}{1-2\lambda_0+2\lambda_{0\xi}}\right)}(2+\log{(1-2\lambda_0+\lambda_{0\xi})}+\log{(1-2\lambda_0+2\lambda_{0\xi})})\right)\\
		+&\frac{A_{1}\beta_1^{(n_f)}}{4\pi\beta_0^{(n_f)3}}\frac{2\lambda_\xi+\log{(1-\lambda_\xi)}(2+(1-\lambda_\xi)\log{(1-\lambda_\xi)})}{1-\lambda_\xi}+\frac{B_{1}}{2\pi\beta_0^{(n_f)}}\log{(1-\lambda_\xi)}\\
		+&\frac{H_{1}}{2\pi\beta_0^{(n_l)}}\log{(1-2\lambda_0+2\lambda_{0\xi})}.
	\end{split}
\end{equation}

\bibliography{references}

\end{document}